\documentclass[11pt]{iopart}
\usepackage{hyperref}
\usepackage[margin=0.7in]{geometry}
\expandafter\let\csname equation*\endcsname\relax
\expandafter\let\csname endequation*\endcsname\relax
\usepackage{xcolor}
\usepackage{graphicx}
\usepackage{dcolumn}
\usepackage{subcaption}
\usepackage{bm}
\usepackage{mathtools}
\usepackage{soul}
\usepackage[font=small]{caption}
\usepackage{float}
\usepackage{MnSymbol}
\usepackage{wasysym}
\usepackage{braket}
\usepackage{nicefrac}

\definecolor{color0}{rgb}{0.886274509803922,0.290196078431373,0.2}
\definecolor{color1}{rgb}{0.203921568627451,0.541176470588235,0.741176470588235}
\definecolor{color2}{rgb}{0.596078431372549,0.556862745098039,0.835294117647059}
\definecolor{color3}{rgb}{0.984313725490196,0.756862745098039,0.368627450980392}
\definecolor{color4}{rgb}{0.556862745098039,0.729411764705882,0.258823529411765}
\definecolor{color5}{rgb}{1,0.709803921568627,0.72156862745098}
\definecolor{color6}{rgb}{0,0.75,0.75}

\begin{document}

\title[New procedure to estimate plasma parameters through the q-Weibull distribution....]{
New procedure to estimate plasma parameters through the q-Weibull distribution by using a Langmuir probe  in a cold plasma }

\author{F. J. Gonzalez$^{[1,2]}$, J. I. Gonzalez$^{[1]}$, S. Soler$^{[3]}$, C. E. Repetto$^{[1,2]}$, B. J. G\'omez$^{[1,2]}$, D. B. Berdichevsky$^{[2,4]}$}
\address{ $^{[1]}$ Facultad de Ciencias Exactas, Ingenier\'{\i}a y Agrimensura (UNR), Av. Pellegrini 250, S2000BTP Rosario, Argentina.}
\address{ $^{[2]}$ Instituto de F\'{\i}sica Rosario (CONICET-UNR), Bv. 27 de Febrero 210 Bis, S2000EZP Rosario, Argentina.}
\address{ $^{[3]}$ Consejo Nacional de Investigaciones Cient\'{\i}ficas y T\'ecnicas (CONICET), Argentina}
\address{ $^{[4]}$ NASA/GSFC, Division 672, Greenbelt 20771, USA}
\vspace{0.2cm}
\address{\textbf{E-mail:} \textbf{\textcolor{blue}{fgonzalez@ifir-conicet.gov.ar}}, \textbf{\textcolor{blue}{bgomez@ifir-conicet.gov.ar}}}
\vspace{0.2cm}
\address{\textbf{Keywords:} cold plasma, q-Weibull distribution, Tsallis statistics}
\vspace{0.2cm}
\address{Plasma Res. Express 4 (2022) 015003: \url{https://doi.org/10.1088/2516-1067/ac4f35}}
\date{\today}

\begin{abstract}
We describe a procedure to obtain the plasma parameters from the {\bf I-V} Langmuir curve by using the Druyvesteyn equation. We propose to include two new parameters, $q$ and $r$, to the usual plasma parameters: plasma potential ($V_p$), floating potential ($V_f$), electron density ($n$), and electron temperature ($T$). These new parameters can be particularly useful to represent non-Maxwellian distributions. The procedure is based on the fit of the {\bf I-V} Langmuir curve with the $q$-Weibull distribution function, and is motivated by recent works which use the $q$-exponential distribution function derived from Tsallis statistics. We obtain the usual plasma parameters employing three techniques: the numerical differentiation using Savitzky Golay (SG) filters, the $q$-exponential distribution function, and the $q$-Weibull distribution function. We explain the limitations of the $q$-exponential function, where the experimental data $V>V_p$ needs to be trimmed beforehand, and this results in a lower accuracy compared to the numerical differentiation with SG. To overcome this difficulty, the $q$-Weibull function is introduced as a natural generalization to the $q$-exponential distribution, and it has greater flexibility in order to represent the concavity change around $V_p$. We apply this procedure to analyze the measurements corresponding to a nitrogen $N_2$ cold plasma obtained by using a single Langmuir probe located at different heights from the cathode. We show that the $q$ parameter has a very stable numerical value with the height. This work may contribute to clarify some advantages and limitations of the use of non-extensive statistics in plasma diagnostics, but the physical interpretation of the non-extensive parameters in plasma physics remains not fully clarified, and requires further research.

\end{abstract}


\maketitle

\section{Introduction}

Langmuir probe is one of the main devices used in plasma diagnostics. For an introduction to Langmuir probe see Refs.~\cite{Cherrington1982,Hershkowitz1989,Merlino2007,Boyd1959,Demidov2002,Abe2013,Bhattarai2017,Woods1994,Knappmiller2006}. The Druyvesteyn method (see Refs.~\cite{Demidov2002,Godyak2011}) makes use of the second derivative of the {\bf {I-V}} measurement, taken with a Langmuir probe, to obtain the electron energy distribution function ({\it{EEDF}}) or the electron energy probability function ({\it{EEPF}}). Druyvesteyn has proved \cite{Druyvesteyn1930} that the second derivative of the probe current {\bf {I(V)}} is related to the {\it{EEDF}}. After Druyvesteyn's work, measurements made with different probes, plasma conditions and positions of the probe inside the plasma chamber have shown that many {\it{EEDF}} are not Maxwellians \cite{Godyak2011,Cherrington1982,Waymouth1989,Hoegy1999,Taccogna2016,Giono2017,Fiebrandt2017}. In this line of development, Bi-Maxwellian, Druyvesteyn and Bi-Druyvesteyn distributions have been proposed to analyze plasma parameters as shown in Refs.~\cite{Tan1973,Godyak1993,Gudmundsson2001,Seo2004,Choe2009,Adams2017,Li2019,Sharma2018}. 

The {\it{EEDF}} obtained by the Druyvesteyn method is valid in general to any convex probe geometry \cite{Druyvesteyn1930}. When {\it{EEDF}} is known, whether it is Maxwellian or not, plasma parameters can be calculated. The plasma parameters are: electron temperature ($T$), electron density ($n$),  plasma potential ($V_p$), and floating potential ($V_f$).

The plasma potential $V_p$ is the potential at the probe location, and it is determined by the probe voltage at the concavity change of the {\bf{I-V}} curve. The floating potential $V_f$ is the potential when the probe is electrically floating, i.e. when the net current collected is zero.

Moreover, we will study a cold plasma, which is defined as a plasma with electron temperature $T<10^5$ K, and we will use pressures around 2 Torr and low enough discharge currents, which implies that the distribution is non-Maxwellian. 

To address non-Maxwellian distribution, we consider the fact that the non-extensive statistics initiated with Tsallis \cite{Tsallis1988,Tsallis1995,Tsallis1997} was proposed to analyze plasma physics and long range interaction problems \cite{Boghosian1996,Anteneodo1997,Silva1998,Lima2000,Du2004,Jiulin2007,Budini2015,Qiu2018,Sun2020,ElBojaddaini2020}. In particular, we study experimental measurements obtained in the plasma cathode sheath, and we rely on Ref.~\cite{Sharifian2014}, because we understand they were the first ones to study plasma sheath with the non-extensive thermodynamics of the electrons. Tsallis statistics considers a non-additive definition of entropy with a real $q$ parameter which represents the degree of non-extensivity, and in the case $q=1$, the entropy definition becomes additive and the old definition from Boltzmann is recovered. In this non-extensive framework, Silva\cite{Silva1998} in 1998 and Qiu and Liu\cite{Qiu2018} in 2018 have proposed their electron probability distributions, these results are summarized in Appendix A. More recently, in 2020, Qiu et al.\cite{Qiu2020} obtained a theoretical derivation of the contribution of electrons to the {\bf {I-V}} Langmuir curve starting from the supposition of Maxwell equilibrium distribution, but with Boltzmann-Gibbs exponential distribution replaced by a $q$-exponential distribution. In this approach, the $q$ value arises as a new plasma parameter related to the degree of non-extensivity of the electrons. 

This derivation is valid up to potentials $V<V_p$ so the experimental data needs to be trimmed beforehand.

In this work we propose that, in order to include the complete experimentally measured {\bf {I-V}} curve in the framework of non-extensive statistics, a new distribution function should be considered.

Starting from experimental measurements made in a cold plasma and with the Langmuir probe situated at positions near the cathode at the plasma sheath, we show that: \textbf{1.} the Savitzky-Golay method is highly sensitive to the filter parameters and these parameters are not uniquely determined; \textbf{2.} by the fact that expressions based on the $q$-exponential distribution, in particular the one of Ref.~\cite{Qiu2020}, only considers a fragment of the Langmuir curve, specifically only the contribution of the electrons with $V<V_p$, it yields to a worsening of the estimated values of plasma parameters compared to the classical Druyvesteyn method using Savitzky-Golay filter when this expression is applied to the complete Langmuir curve; \textbf{3.} the $q$-Weibull distribution introduced in this work can reproduce the complete experimentally measured I-V curve. 

The $q$-Weibull cumulative distribution function (cdf) is not only proposed because of the aforementioned, it is also proposed because it is the natural generalization of other proposed {\it{EEPFs}}. This is discussed in Sec. \ref{sec:qweibull} and Appendixes B and C.

This work is organized as follows. Section~\ref{sec:experimental} provides a description of the experimental set-up. Section ~\ref{sec:druyvesteyn} briefly discusses the Druyvesteyn method. Then, we consider three different approaches to analyze the {\bf {I-V}} curves by using the Druyvesteyn method: (i) in Sec.~\ref{sec:SG}, the usual filtering and numerically differentiating of the data; (ii) in Sec.~\ref{sec:qiu}, the non-extensive expression from Qiu et al. \cite{Qiu2020} ; (iii) in Sec.~\ref{sec:qweibull}, the $q$-Weibull cumulative distribution. Section~\ref{sec:discussion} is devoted to the discussion, and conclusions are summarized in Sec.~\ref{sec:conclusion}.

\section{Experimental set-up}\label{sec:experimental}

\subsection{Plasma Generation}\label{sec:plasmageneration}

In the present work, the characterization of a nitrogen plasma (N$_2$) at (2.00 $\pm$ 0.01) Torr (typical value used i.e., in surface treatments) was performed in a stainless steel chamber, powered by a DC potential of (500$\pm$2) V and with a discharge current of (0.455$\, \pm \, $0.005) A. 

\begin{figure}[htpb]
    \centering
    \includegraphics[width=0.3\linewidth]{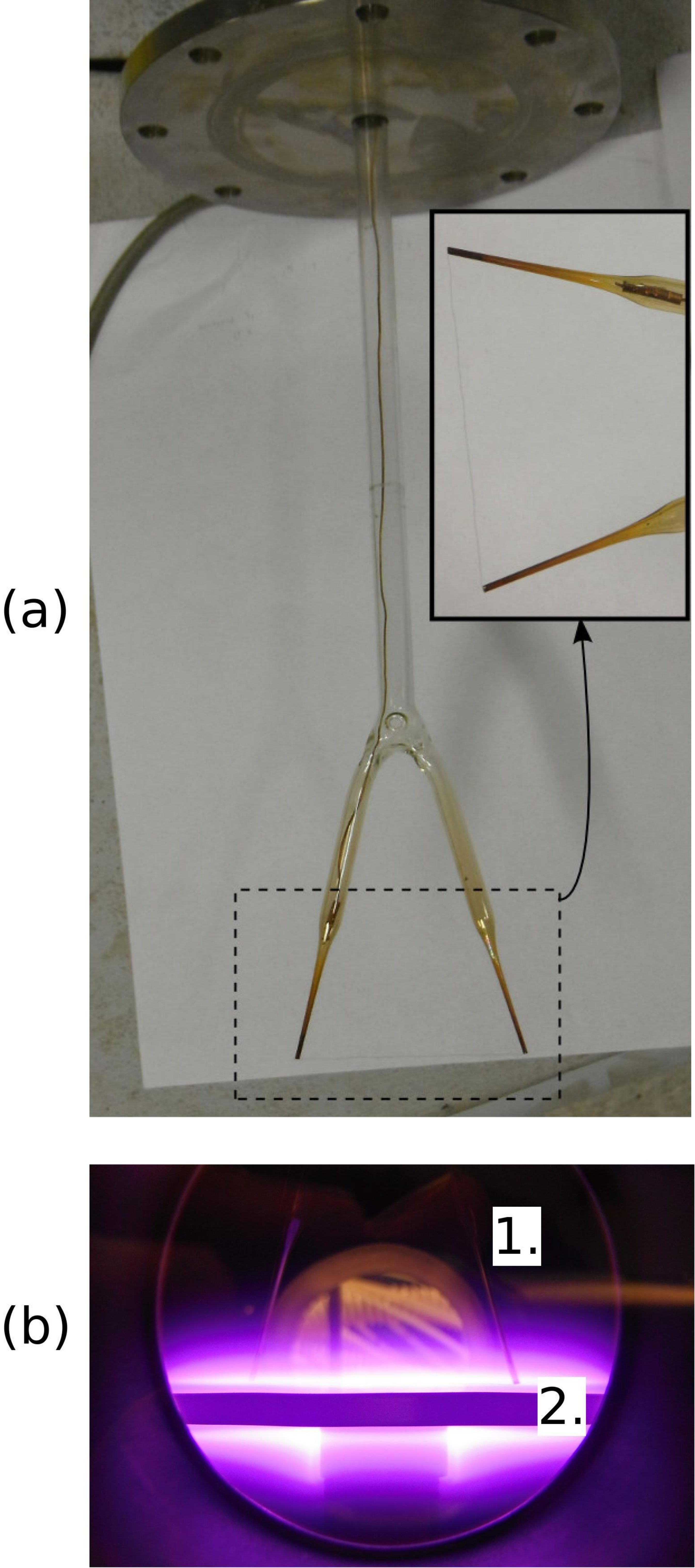}
    \caption{Langmuir probe used in the {\bf {I-V}} measurements. \textbf{(a)} Probe with a tungsten wire. \textbf{(b)} Probe located at millimeters from the cathode in the plasma sheath: \textbf{1.} Probe, \textbf{2.} Cathode.}
    \label{fig:probe}
\end{figure}

\begin{figure}[htpb]
\centering
\includegraphics[width=0.5\linewidth]{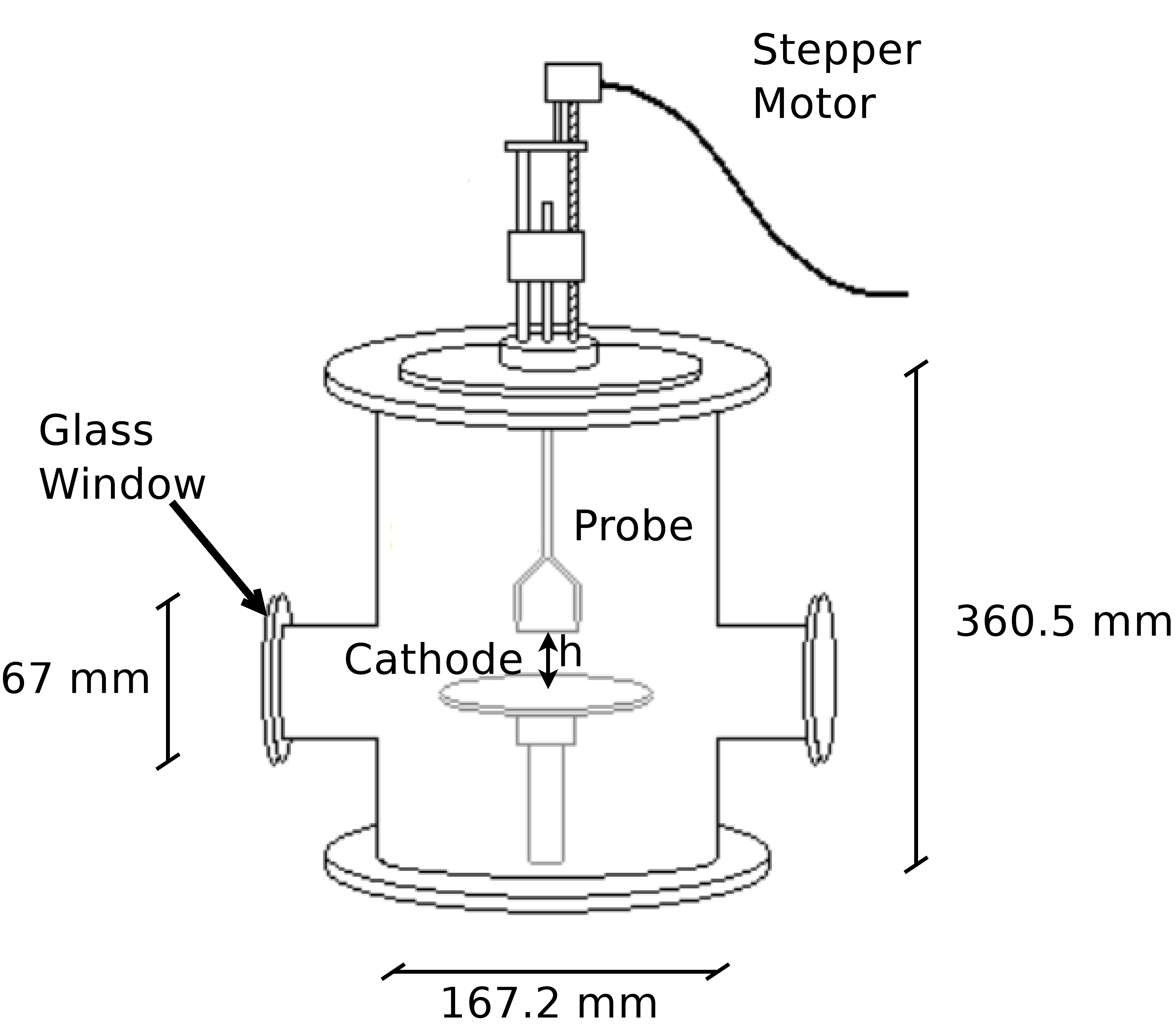}
\caption{Dimensions of the chamber and some components of the experimental set-up.}
\label{fig-diagr-system}
\end{figure}

\subsection{Obtaining the {\bf {I-V}} curve}
\label{sec:ivcurve}

The voltage-current characteristic curve ({\bf {I-V}} curve) was measured with a cylindrical probe constructed with a tungsten wire and a glass support in ``V'' shape (see Figure~\ref{fig:probe}(a)).  The dimensions of the probe are: diameter $d=(30\, \pm \, 1)$ $\mu$m and length $l=(6.40 \,\pm \,0.02)$ cm. We decided to use this relatively long probe only due to technical reasons, and other probes could also be used. 
The system under study consists of both a plasma and a probe system. The last one includes a specific probe and the measurement equipment. We have not studied in detail the influence of the probe system in the plasma conditions, and probably with another probe system the obtained curves could have been different. However, this work aims to describe a procedure to use the experimental data from the {\bf {I-V}} curve to obtain a precise estimation of the plasma parameters of the whole system, and this procedure is planned to be applied with different probe systems.

The probe was placed at different heights ($h$) starting from the cathode, in order to obtain measurements in the zones of the  cathode glow and the cathode dark space. The dimensions of the chamber and the height $h$ are shown in Fig. \ref{fig-diagr-system}. A total of eight heights were considered at distances of 2.2, 3.4, 4.3, 5.1, 6.2, 7.3, 8.2 and 9.4 mm from the cathode. These positions were set by a stepper motor, which is able to move the probe vertically with enough precision.

\begin{figure}[H]
\centering
\includegraphics[width=0.95\linewidth]{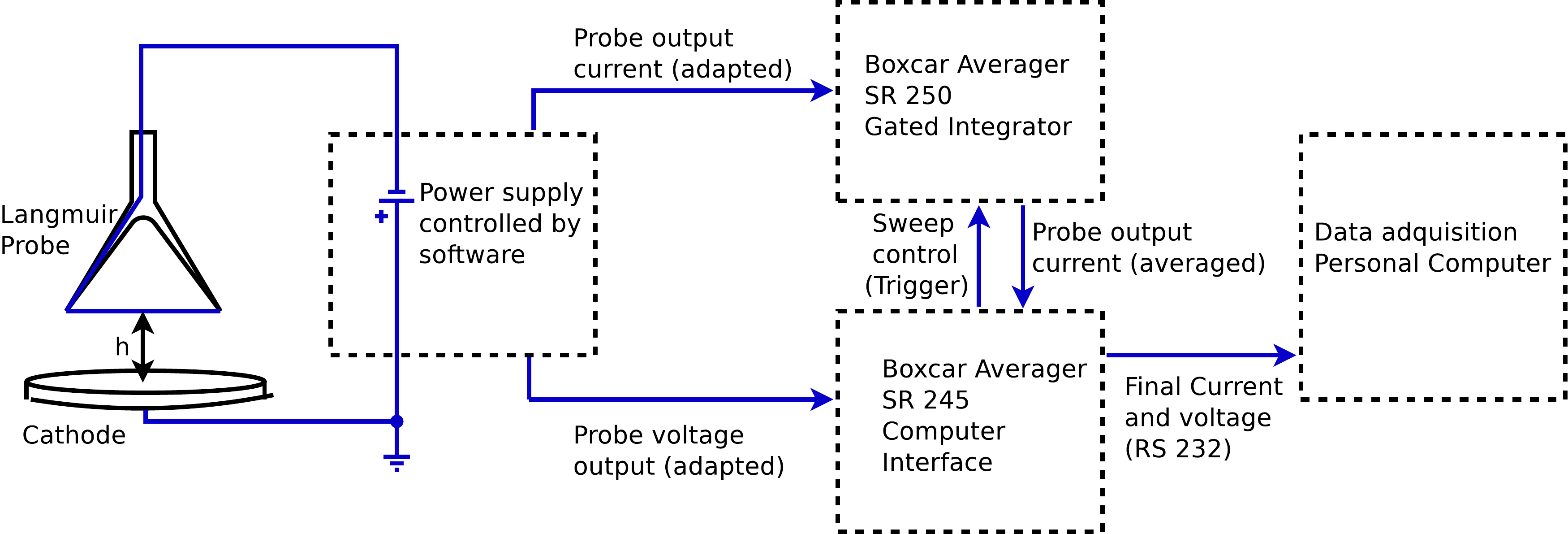}
\caption{Diagram of the probe polarization and the measurement equipment.}
\label{fig-diagr-probe}
\end{figure}

A diagram of the probe measurement set-up is shown in Fig. \ref{fig-diagr-probe}. A power source, which is controlled by software, is connected to the probe. This source provides the current and voltage probe as outputs, i.e. the contribution of both electrons and ions. To measure the current, we use a Boxcar Averager SR 250 Gated Integrator, averaging over 200 measurements of the current in a fixed time window starting from 800$\mu s$ after each voltage step. On the other side, the voltage of the probe and the averaged current are measured by a Boxcar Averager SR 245 Computer Interface. Each {\bf{I-V}} curve consists of a sweep in voltage of 100 points. Further details about the experimental set-up can be found in Ref. \cite{Isola2009}, where a similar configuration was used.

\begin{figure}[H]
\centering
\includegraphics[width=0.7\linewidth]{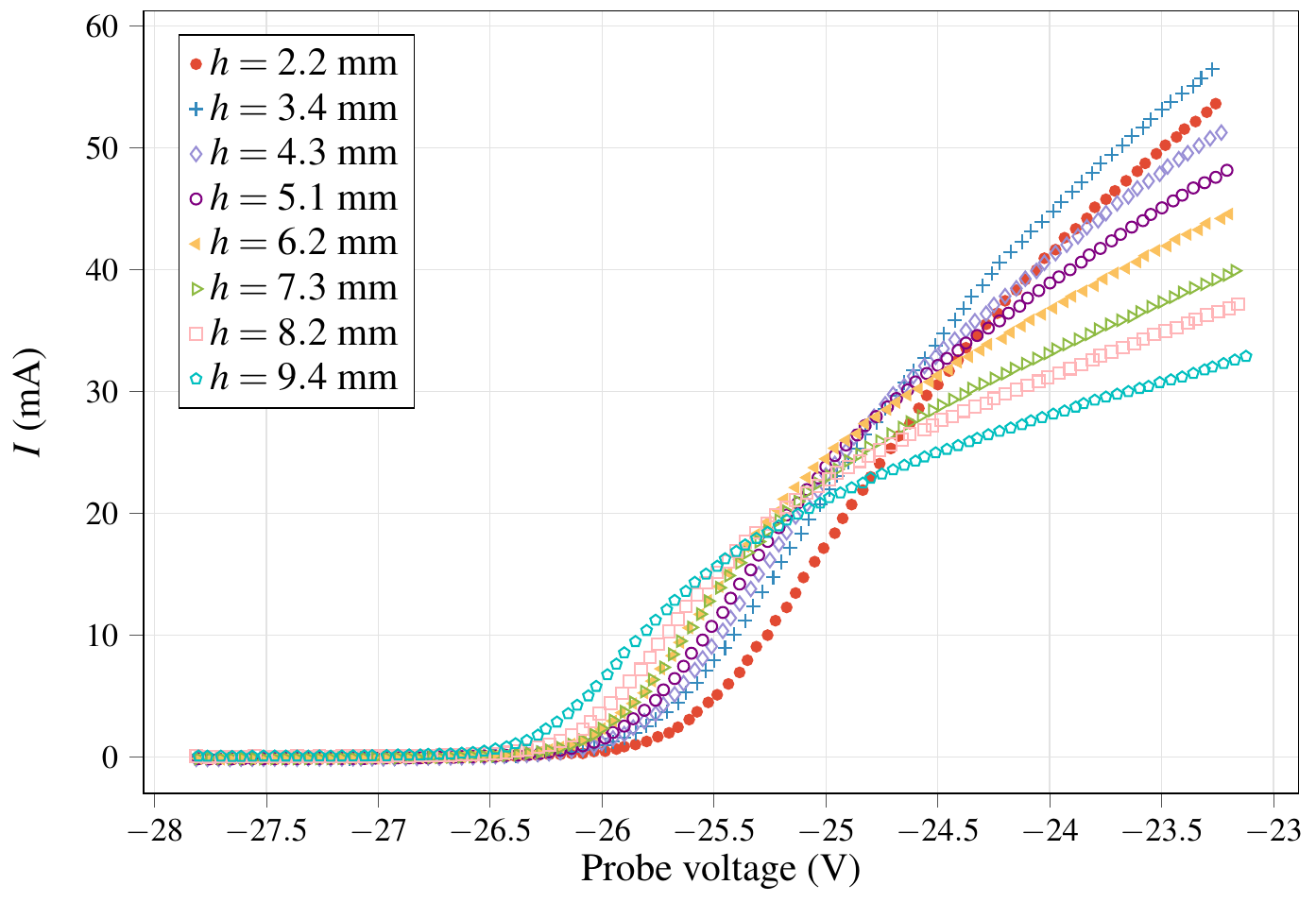}
\caption{{\bf {I-V}} measurements with the probe placed at different distances, $h$, from the cathode.}
\label{fig-iv}
\end{figure}

For each one of the eight heights considered in this work we carried out three sweeps, and we verified that {\bf {I-V}} curves remained constant in time. The measurement results are shown in Fig.~\ref{fig-iv}. 

In this last paragraph, we mention the values of some useful parameters to identify the plasma operating regime for the measurements presented. In a neutral gas with  $p = 2$ Torr and $T = 300$ K, we have $N_g= 6.43 \cdot 10^{22} m^{-3}$ molecules per unit volume, and by using the gas kinetic diameter ($d_g$), whose value for $N_2$ is\cite{Huxley1974,Ismail2015} $d_g$ = $3.64 \cdot 10^{-10}m $, then the electron mean free path ($\ell_m$) can be calculated approximately within the framework of kinetic theory of ionized gases\cite{McDaniel1964,Huxley1974,Biberman1987,Ismail2015} by
\begin{equation}
    \ell_m = \frac{4}{\pi N_g d_g^2} \approx 150 \mu m
    \label{eq:meanfreepath}
\end{equation}
The effective dimension of the cylindrical probe, $a$, can be defined by\cite{Demidov2002}
\begin{equation}
    a = \frac{d}{2}\ln{ \left(\frac{\pi l}{2d} \right)}\approx 120\,\mu{\rm m}
    \label{eq:probedimension}
\end{equation}
where $l$ and $d$ are the length and diameter of the probe, it is verified that $\ell_m > a$. Therefore, we conclude that the plasma is in the collision-free regime\cite{Demidov2002,Hershkowitz1989}.
The Debye length ($\lambda_D$) can be calculated by
\begin{equation}
    \lambda_{D}=\sqrt{\varepsilon_0 k_B T / e^{2}n } 
    \label{eq:debyelength}
\end{equation} 
where $k_B$ is the Boltzmann constant and $e$ the electron charge. Using the values of $T$ and $n$ estimated in this work, we obtain $\lambda_{D}$ in the range of $15-45\,\mu{\rm m}$. The plasma sheath thickness, $h_s$, is considered\cite{Faudot2019,Demidov2002} from one up to five times greater than the Debye length, finally it is verified that $\ell_m > a\gtrapprox h_s$ which implies the plasma is in the conventional thin sheath subregime\cite{Demidov2002}.

\section{Druyvesteyn method}\label{sec:druyvesteyn}
By using the Langmuir expression for the electron probe current \cite{Langmuir1926}, Druyvesteyn\cite{Druyvesteyn1930} demonstrated that the current of a probe ({\bf {I(V)}}) allows determining the {\it{EEDF}}. Such relation involves the second derivative $d^{2}I_{e,ret}/dV^{2} =I_{e,ret}^{\prime\prime}$ of the retarded electron current with $V<V_p$. The probe current {\bf {I(V)}} is the sum of both electron retarding current $I_{e,ret}$ ($V<V_p$) and ion saturation current $I_{i,sat}$ ($V<V_p$). But if the probe bias is not too negative, it is verified\cite{Godyak1992RF} $I^{\prime\prime}=I_{e,ret}^{\prime\prime}+I_{i,sat}^{\prime\prime}\approx I_{e,ret}^{\prime\prime}$. This approximation has been analyzed in detail in some works, for example in Refs. \cite{Godyak1992RF,Godyak2011}. Therefore, in the following equations we use $I$ instead of $I_{e,ret}$.
Finally, we consider the total current $I^{\prime\prime}$ restricted to the values $V<V_p$. Considering the works in Refs.~\cite{Demidov2002,Godyak2011,Lieberman}, in an isotropic plasma, the relation between $I^{\prime\prime}$ and the {\it{EEDF}} ($F(\varepsilon)$), with $\varepsilon=eV^{\prime}$ is provided by the Druyvesteyn formula 
\begin{equation}
    F(eV^{\prime})=\frac{4}{e^3S}\sqrt{\frac{emV^{\prime}}{2}} \left. \frac{d^{2}I}{dV^{2}}\right| _{V=V_p-V^{\prime}} \; \; , \; \; V^{\prime}=V_p-V > 0
    \label{ec:DruyvesteynF}
\end{equation}
where $e$ is the electron charge in absolute value, $S$ is the probe area, and $V^{\prime}=V_{p}-V$ is a variable change from $V$ to $V^{\prime}$, where $V_{p}$ is the plasma potential defined as the value of $V$ where $I^{\prime \prime}$ vanishes. Physically, $F(\varepsilon)=F(eV^{\prime})$ represents the number of electrons in the volume element having energies between $\varepsilon$ and $\varepsilon+d\varepsilon$.

Instead of the {\it{EEDF}} ($F(\varepsilon)$), the {\it{EEPF}} ($f(\varepsilon)$) can be considered, and the following relationship is verified in isotropic plasmas\cite{Godyak2011}
\begin{equation}
    F(\varepsilon)=\sqrt{\varepsilon}f(\varepsilon)\,.
    \label{ec:EEDF-EEPF}
\end{equation}
The {\it{EEPF}} has the advantage of being proportional to $d^2I/dV^2$ as it can be seen by using Eq.(\ref{ec:EEDF-EEPF}) in Eq.(\ref{ec:DruyvesteynF}) 
\begin{equation}
   f(eV^\prime)=\frac{4}{e^3S}\sqrt{\frac{m}{2}}\left. \frac{d^2I}{dV^2}\right| _{V=V_p-V^{\prime}} \; , \; V^\prime=V_p-V > 0 \, .
    \label{ec:Druyvesteynf}
\end{equation}

Knowledge of the {\it{EEDF}} or {\it{EEPF}} allows the calculation of the electron density $n$ and the electron temperature $T$ (defined in general to a non-Maxwellian {\it{EEDF}} as the measure of the average electron energy $\braket{\varepsilon}$ by the relationship $\braket{\varepsilon}=\nicefrac{3}{2}n k_B T$) 
\begin{equation}
    n=\int_0^\infty \!\!F(\varepsilon) d\varepsilon = \int_0^\infty \!\! \sqrt{\varepsilon} f(\varepsilon) d\varepsilon \,,
    \label{ec:N}
\end{equation}
\begin{equation}
    T=\frac{2}{3}\frac{1}{nk_B} \int_0^\infty \!\!\varepsilon F(\varepsilon) d\varepsilon =\frac{2}{3}\frac{1}{nk_B} \int_0^\infty \!\! \varepsilon^{3/2} f(\varepsilon) d\varepsilon \,.
    \label{ec:Te}
\end{equation}

By using Eq.(8), we obtain an uniquely defined temperature which corresponds to the temperature obtained directly from the {\bf{I-V}} curve. This calculated value corresponds to the Maxwellian concept of temperature only in the case of a Maxwellian EEDF. The same can be said for the electron density and plasma parameters in general, where the parameters are unique and defined from the Langmuir \textbf{I-V} curve.

In summary, to obtain the {\it{EEPF}} we make use of Eq.~(\ref{ec:Druyvesteynf}) where the change of variable $V=V_p-V^\prime$ involves the inversion and shifting of the voltage axis of the original {\bf{I-V}} curve. The {\it{EEPF}} is usually represented in semilog-scale, where a linear behavior is characteristic of a Maxwellian plasma\cite{Godyak1990} and a nonlinear behavior is characteristic of a non-Maxwellian plasma. 

\section{Data analysis}\label{sec:dataanalysis}

\subsection{Using numerical differentiation: Savitzky-Golay filter}\label{sec:SG}

The second derivative of the curve $d^{2}I/dV^{2} =I^{\prime\prime}$ is usually obtained by the procedure of filtering and numerically differentiating the {\textbf{I-V}} curve.

There are many possibilities to obtain this second derivative by using different techniques, ranging from analog solutions (e.g. an analog electronic circuit) to digital smoothing (e.g. a digital electronic circuit or a post-processing of the measured data in a software). In the digital smoothing technique, many alternatives are also available, the more commonly used\cite{Magnus2008} are: Savitzky–Golay filter, the Gaussian filter, polynomial fitting and the Blackman filter. In each of these possibilities, some parameters are needed and may not be obvious the optimal election of them. In this work in particular, we only use the  Savitzky-Golay\cite{Savitzky-Golay,Roh2015} filter (SG), and its parameters are: {\textbf{(M)}} the degree of the polynomial and {\textbf{(n)}} the number of points (frame length).

\begin{figure}[H]
\centering
\includegraphics[width=0.6\linewidth,trim=2.5cm 14.5cm 8.5cm 1.8cm]{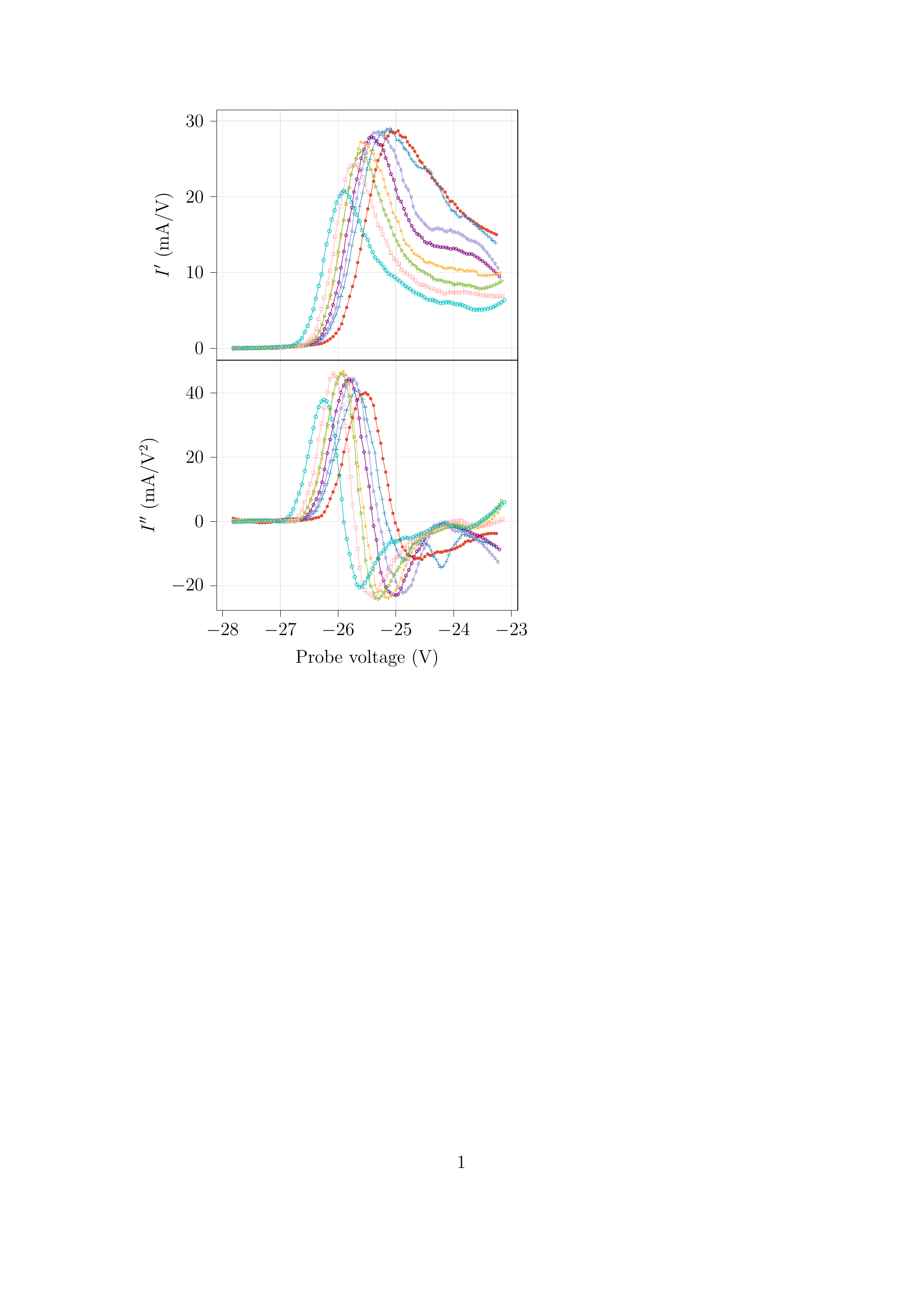}
\caption{ $I^\prime(V)$ and $I^{\prime \prime}(V)$ curves obtained from experimental data using the Savitzky-Golay filter with the probe at different heights measured from the cathode: \textcolor{color6}{$\pentagon$} 9.4, \textcolor{color5}{\large$\square$} 8.2, \textcolor{color4}{$\triangleright$} 7.3, \textcolor{color3}{$\blacktriangle$} 6.2, \textcolor{violet}{\large$\circ$} 5.1, \textcolor{color2}{\small$\lozenge$} 4.3, \textcolor{color1}{\large$\plus$} 3.4 and \textcolor{color0}{\small$\bullet$} 2.2 mm. The solid lines between symbols serve only as a guide. The upper subplot has been calculated by using a one-step application of a SG onefold differentiation, and the lower subplot by a one-step application of a SG twofold differentiation. The upper subplot is not used in the calculation of plasma parameters and is plotted only as a graphical visualization.}
\label{fig-iv-Druv}
\end{figure}

Note that these numerical procedures involve all the measured data, and only after obtaining the second derivative, the restriction $V<V_p$ (see Sec. \ref{sec:druyvesteyn}) is applied. This is one of the reasons of the effectiveness of the procedure.
Figure~\ref{fig-iv-Druv} shows the smoothed  $I^\prime$ and $I^{\prime \prime}$ curves obtained from experimental data by using SG. 
 It is important to highlight again that there are many possibilities to obtain the second derivative using SG, and these different options lead to different plasma parameters. In the following, we consider:

{\renewcommand\labelitemi{}
\begin{itemize}
    \item\textbf{(1.)} a SG smoothing of the {\bf{I-V}} curve and a numerical differentiation to obtain the first derivative, then a new SG smoothing and a numerical differentiation to obtain the second derivative 
    \item\textbf{(2.)} a two-step application of a SG onefold differentiation
    \item\textbf{(3.)} a one-step application of a SG twofold differentiation
    \item\textbf{(4.)} a SG smoothing of the {\bf{I-V}}  curve and an application of \textbf{(2.)} or \textbf{(3.)}
    \item\textbf{(5.)} a use of \textbf{(2.)} with an intermediate SG smoothing between the steps
\end{itemize}}

We have implemented all these options also using different values to the parameters \textbf{M} and \textbf{n}. To select the optimal values to \textbf{M} and \textbf{n} we have analyzed the final smoothness of the second derivative, also the existence of values $I^{\prime \prime}<0$ which are undesirable (because the {\it{EEPF}} in semilog-scale would have undefined values), and also the plasma parameters $T$ and $n$ obtained from the {\it{EEPF}}. According to our measurements, we have found that options \textbf{(1.)} and \textbf{(2.)} provide similar results of $T$ and $n$ plasma parameters, with the optimal values \textbf{M}=2 and \textbf{n}=19 for the first derivative and \textbf{M}=2 and \textbf{n}=17 for the second derivative.

\begin{figure}[H]
\centering
\includegraphics[width=0.75\linewidth,trim=2.5cm 14.5cm 8.5cm 1.8cm]{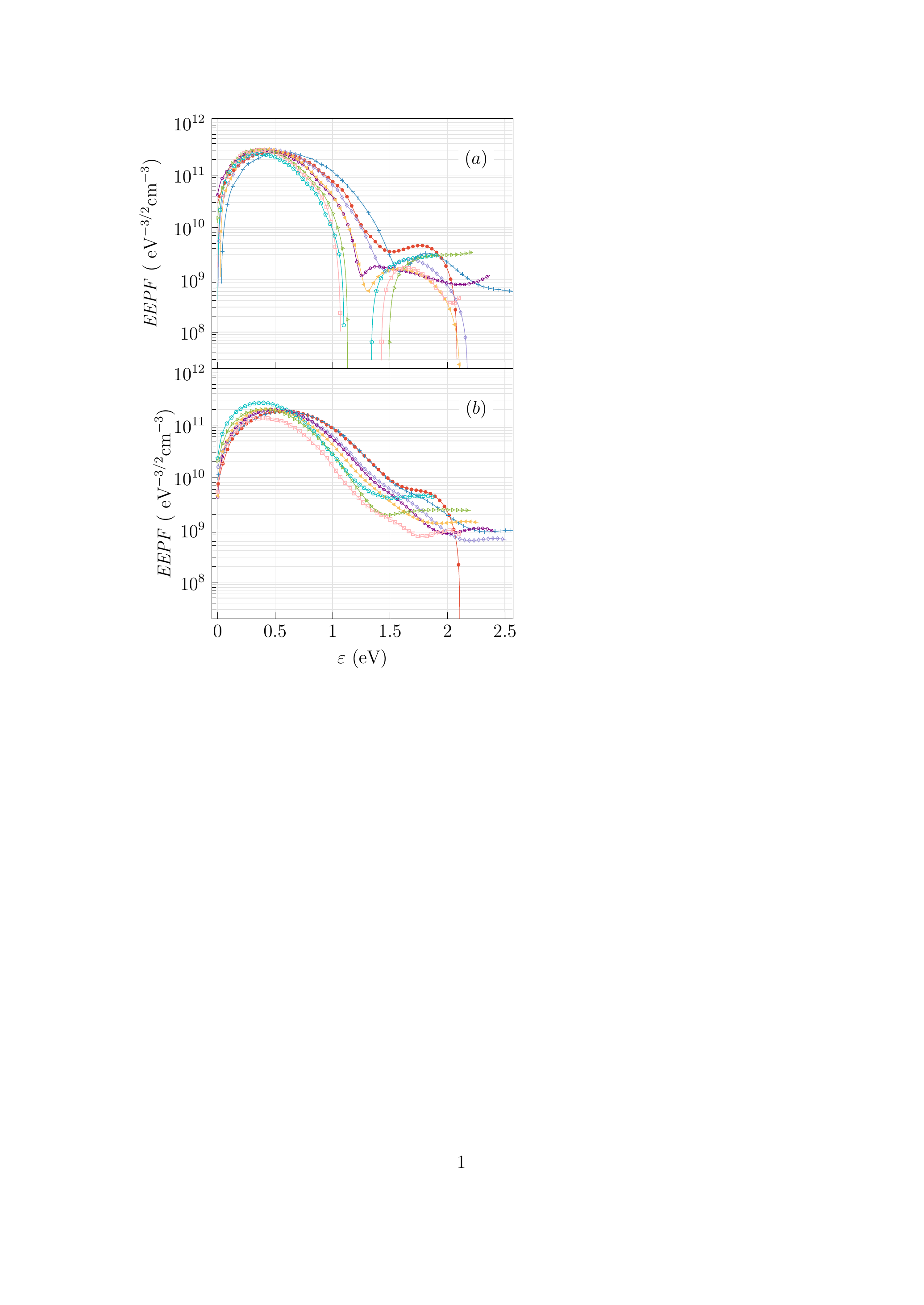}
\caption{($a$) {\it{EEPF}} curves obtained from $I^{\prime\prime}(V)$ using the two-step application of a Savitzky-Golay onefold differentiating filter with the experimental data and Druyvesteyn equation, for the probe at different heights measured from the cathode: \textcolor{color6}{$\pentagon$} 9.4, \textcolor{color5}{\large$\square$} 8.2, \textcolor{color4}{$\triangleright$} 7.3, \textcolor{color3}{$\blacktriangle$} 6.2, \textcolor{violet}{\large$\circ$} 5.1, \textcolor{color2}{\small$\lozenge$} 4.3, \textcolor{color1}{\large$\plus$} 3.4 and \textcolor{color0}{\small$\bullet$} 2.2 mm. The solid lines between symbols serve only as a guide. ($b$) {\it{EEPF}} curves obtained from $I^{\prime\prime}(V)$ using a one-step application of a SG twofold differentiating filter for the same  experimental data.}
\label{fig:EEPF2xDruv}
\end{figure}

Figure~\ref{fig:EEPF2xDruv}(a) shows the {\it{EEPF}} calculated with Eq.~(\ref{ec:Druyvesteynf}) from $I^{\prime \prime}$ to the case \textbf{(2.)}. There are some curves that show a ``gap'' between 1 and 1.5 eV, these are numerical artifacts.

To avoid such ``gap'' we have tried options \textbf{(4.)} and \textbf{(5.)} but without favourable results. Finally, we implemented option \textbf{(3.)} with \textbf{M}=2 and \textbf{n}=19 and that fixed the ``gap'' problem. Figure~\ref{fig:EEPF2xDruv}(b) shows the {\it{EEPF}} of the last case.

Now that we have solved the ``gap'' problem, there is another important point to consider: the variations of plasma parameters, principally $T$, with the different options mentioned. We have observed that $T$ is very sensitive to the options considered and even to the specific filter parameters chosen, to be more specific we have found that the different options and filter parameters affect the estimation of the plasma potential $V_p$ and this is traduced in a high variation of the T value. For example with option \textbf{(2.)} to the height $h$=2.2 mm, we obtain $V_p$=-25.06 V and T=4900 K, and with option \textbf{(3.)} to the same height we obtain $V_p$=-25.01 V and T=5750 K. Moreover, if we set $V_p$=-25.01 V (obtained from \textbf{(3.)}) and use this value with option \textbf{(2.)} we obtain T=5300 K, so this result shows the importance of the $V_p$ value. 
In the following we consider only option \textbf{(3.)} which does not present the ``gap''. We evaluate the plasma parameters using the $V_p$ value calculated in two different ways: from the maximum value of $I^\prime$, and from $I^{\prime \prime}=0$. The $V_p$ value represents the starting potential which sets the 0 $eV$ in the EEPF, and it is also the inferior limit of the integration in Eqs.~(\ref{ec:N}) and (\ref{ec:Te}). When we use the criteria of $V_p$ defined as the maximum value of $I^\prime$, it is found a more precise estimation of $V_p$ because it requires only a unique numerical differentiation. However, the plasma parameters are obtained from the integration of the second derivative $I^{\prime\prime}$, so if we use the $V_p$ value obtained from the first derivative to integrate the second derivative we find an inconsistency in the chosen $V_p$ potential. Therefore, we consider it is better to use the second criteria, i.e. obtaining $V_p$ from $I^{\prime \prime}=0$, because we are being consistent both with the inferior limit of integration chosen and the argument of integration.
Finally, the values $V_p$, $V_f$, $n$, and $T$ (using option \textbf{(3.)}) are summarized in Table~\ref{tab:druv}, for each height considered. The floating potential $V_f$ is calculated from $I(V_f)=0$ by the interpolation of the {\bf{I-V}}  curves. The uncertainties were obtained taking into account the sensitivity of the results observed using different SG parameters. Specifically, we changed the value of the frame length filter parameter (\textbf{n}), which is \textbf{n}=19 in the optimal case, to \textbf{n}=17 and \textbf{n}=21. Then we evaluate $V_p$, $n$ and $T$, and as a result we obtain a bound in the uncertainty of these parameters.

\begin{table}[H]
\caption{Plasma parameters obtained using the Savitzky-Golay filter.}
\label{tab:druv}
\centering
\begin{tabular}{ c  c  c  c  c }
\br
$h\,\pm \! 0.1$ & $V_p\, \pm \! 0.05$ & $V_f\, \pm \! 0.05$ & $n\, \pm \!0.1$ & $T\,\pm\! 600$ \\ (mm) & (V) & (V)  & ($10^{11}$/cm$^{3}$) & (K)  \\ [0.5ex]
\mr
2.2 &  -25.01 & -26.58 & 1.6 & 5700 \\
3.4 & -25.16 & -26.53 & 1.5 & 5400 \\
4.3 & -25.30 & -26.83 & 1.4 & 4600 \\ 
5.1  &  -25.39 & -26.81 & 1.4 & 4700 \\ 
6.2  &  -25.53 & -26.71 & 1.3 & 4400 \\ 
7.3  & -25.60 & -26.73 & 1.1 & 3600 \\ 
8.2  & -25.70 & -26.81 & 1.0 & 4000 \\
9.4  & -25.90 & -26.96 & 0.9 & 3700 \\
\br
\end{tabular}
\end{table} 

A last comment can be made concerning the depleted shape of the EEPFs in Fig.~\ref{fig:EEPF2xDruv} for energies below 0.5 eV. 

Below, we analyze our particular measurements with Nitrogen at a relatively high pressure (2 Torr) and a relatively low discharge current (0.455 A). We define $\Delta$ as the distance between $0$ eV to the maximum of the low energy peak, as it can be seen in Fig.~\ref{fig:EEPF2xDruv} we obtain $\Delta \approx ( 0.3 - 0.6 ) eV$. From practical considerations as it is depicted in  Ref.~\cite{Godyak1992RF} it is useful to compare $\Delta$ with the electron temperature obtained from the \textbf{I-V} curves, and a ``good quality experiment'' should satisfy $\Delta < e T$. In Table~\ref{tab:druv}, we obtain in the case of $T=5700$ K $\rightarrow \Delta =0.6 eV > eT = 0.49 eV$, and for the lower temperature $T=3700$ K $\rightarrow \Delta =0.3 eV < eT = 0.32 eV$, which implies we are at the limit of ``good/not good quality experiment'' based on this criterion. This depletion at low electron energies has been attributed in Ref.~\cite{Godyak1992RF} to probe contamination and low input impedance in the probe system, and plasma density depletion around large probes and high gas pressures. In the case of our measurements, and based on Ref.~\cite{Godyak2021}, we consider that probably the predominant effects related to low energy depletion come from the large probe used ($L=6.4$ cm), which can cause an unbalanced electron energy and charge current distribution along the probe, from the high pressure which incentivizes the EEPF to have a more convex shape or to be Druyvesteyn-like, and from the nearness of the probe with the cathode which can favor the low energy electrons to accelerate and also the covering of the probe with the material evaporated by the cathode. Even if we figure out the precise physical mechanism of this depletion, it is known from Ref.\cite{Godyak1992RF} that a small ``residual'' value of $\Delta$ less than $0.5$ eV can not be avoided, probably as a consequence of electron reflection, secondary-electron emission, inhomogeneity of the probe work function along the probe collecting surface and the convolution effect usually accompanying differentiation. For practical considerations, it is common to extrapolate this low electron energy zone to obtain more precise values of the plasma parameters, as it is shown in Refs.~ \cite{Godyak1992RF} and \cite{Godyak1990PhRevLett}. However, in this work, in view of the complicated and a priory unknown mechanisms which are involved, we decided to use the experimental \textbf{I-V} curves without modifications. We will obtain plasma parameters which will include the probe and plasma systems as a unique system, with all the previously mentioned effects included in the calculation. It is important to remark that with this procedure we will obtain different plasma parameters when different probe systems are used, and a more precise knowledge of the involved mechanism would allow us, in a future calculation, to correct the plasma parameters in each experimental set-up. This alternative point of view opens the door to studying plasma parameters with homemade probes and avoiding the use of sophisticated probe systems.

\subsection{Using a $q$-exponential based expression}\label{sec:qiu}

Recently, in 2020, Qiu et al. (Eq.(1) in Ref.~\cite{Qiu2020}) obtained an expression for the current from the probe using the theoretical framework of non extensive $q$-statistics. They considered plasma behavior as Maxwellian and replaced the Boltzmann-Gibbs exponential distribution function by the $q$-exponential function; this yields an expression for the current where the $q$ value arises as a new plasma parameter. Based on Ref.~\cite{Qiu2020} we propose to fit the experimental data with the function

\begin{equation}
    I_Q(V;q^*,A,B)=A\left(  C_{q^*}  \left( 1+ (q^*-1) B (V-V_p) \right)^{\frac{1}{q^*-1}+\frac{1}{2}} - ( 1-(q^*-1)\frac{1}{2}  )^{\frac{1}{q^*-1}+\frac{1}{2}} \right) \; ,
    \label{ec:qiu}
\end{equation}
where the fitting parameters are $q^*$, $A$ and $B$.  The variable $C_{q^*}$ depends on $q^*$ by
\begin{equation}
  C_{q^*}=  \frac{A_{q^*}}{q^*} \sqrt{\frac{m_i}{2\pi m_e}} 
  \label{eq:cq*}
\end{equation}
where $m_e$ and $m_i$ are the electron and ion mass, and $A_{q^*}$ is a $q^*$ dependent normalization parameter whose value is
\begin{equation}
    A_{q^*}=\begin{cases}
     \sqrt{1-q^*}\frac{\Gamma(\frac{1}{1-q^*})}{\Gamma(\frac{1}{1-q^*}-\frac{1}{2})} & ,  \; -1< q^* \le 1 \\
     \\
     \frac{1+q^*}{2}\sqrt{q^*-1}\frac{\Gamma(\frac{1}{q^*-1}+\frac{1}{2})}{\Gamma(\frac{1}{q^*-1})} & , \; q \ge 1
    \end{cases}
    \label{eq:aq*}
\end{equation}

The first term in Eq.~(\ref{ec:qiu}) corresponds to the electron contribution and the second one to the ion contribution. In this equation, the last term is assumed to be a constant in respect to $V$, and it corresponds to the ion saturation current $I_{sat}$. In the notation of Ref.~\cite{Qiu2020} the $A$ and $B$ fitting parameters are given by
\begin{eqnarray}
  \begin{cases} 
  A=e n_e A_p \sqrt{\frac{k_{B}T_{e}}{m_{i}}} \\
  B=\frac{e}{k_{B}T_{e}}  \\
  \end{cases}
  \label{eq:qiupar}
\end{eqnarray}

where $A_p$ is the probe area and $e$ is the electron charge.

Equation (\ref{ec:qiu}) is based on the following assumptions (textually cited from Ref. \cite{Qiu2020})
\begin{itemize}
    \item No magnetic field;
    \item The plasma is thin, so the average free path of electrons and ions is much larger than the probe size (collision-free approximation);
    \item The electrons obey the non-extensive distribution, and the ion temperature is very low ($T_i \approx 0$, cold plasma approximation);
    \item The thickness of the formed sheath is much smaller than the probe size (order of several Debye length $\lambda_D$), thus allowing the use of plate approximation rather than relying on the geometry of the probe\cite{Lipschultz1986};
    \item  The emission of the secondary electron on the surface of the probe can be ignored, and the charged particles that hit on the probe do not react with the probe, namely the probe is an ideal absorber of charged particles.
\end{itemize}
Based on the measurements from this work presented in Sec. \ref{sec:ivcurve}, we consider that all the items are satisfied. The fourth item deserves some discussion. As it was analyzed in Sec. \ref{sec:ivcurve}, we have obtained  $a\gtrapprox h_s$, i.e. the thickness is smaller than the probe size. The fourth item requires $a \gg h_s$, therefore, we are aware that the probe geometry can be a subject of future consideration. However, we will consider that Eq. (\ref{ec:qiu}) can be applied with our measurements.

By definition, Eq.~(\ref{ec:qiu}) is valid for $V < V_p$ values. In order to find the plasma parameters using this approach, we have to ignore some part of the experimental data from the data set, specifically the $V > V_p$ values. To estimate the $V_p$ value, we again reside in a numerical procedure such as the one mentioned in Sec. \ref{sec:SG}. This procedure of ignoring some part of the experimental data, as we will see below, conduces to greater errors in the estimated plasma parameters compared to the SG filtering from Sec. \ref{sec:SG}.

To obtain parameters $A$, $B$ and $q^{*}$ from Eq.~(\ref{ec:qiu}), we do the following procedure: \textbf{1.} we use the values of $V_p$ obtained in Sec. \ref{sec:SG}, then \textbf{2.} we plot Eq.~(\ref{ec:qiu}) by evaluating Eqs.~(\ref{eq:qiupar}) with reasonable values on the right side, and compare it with the {\bf{I-V}} curve to find the initial values for the fitting parameters for each {\bf{I-V}} curve, finally \textbf{3.} we use the initial parameters to find the convergence of the fit. 

\begin{figure}[H]
\centering
\includegraphics[width=0.7\linewidth]{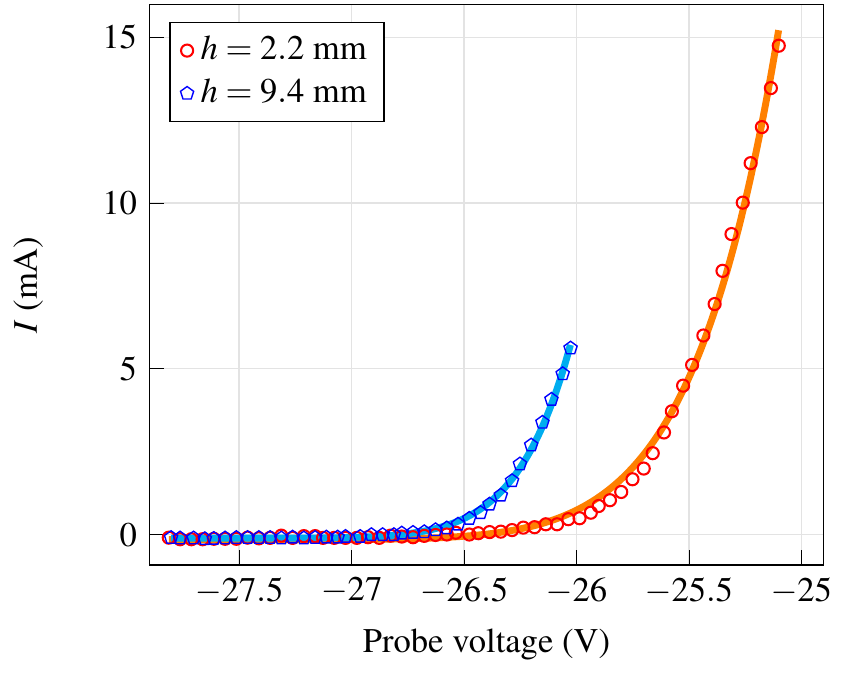}
\caption{(Symbols) Experimental {\bf {I-V}} measurements, with the probe at two different distances from the cathode. (Continuous lines) Fit of the experimental {\bf {I-V}} data using the function defined in Eq.~(\ref{ec:qiu}).}
\label{fig-VI_Qiu}
\end{figure}

\begin{figure}[H]
\centering
\includegraphics[width=0.75\linewidth,trim=2.5cm 14.5cm 8.5cm 1.8cm]{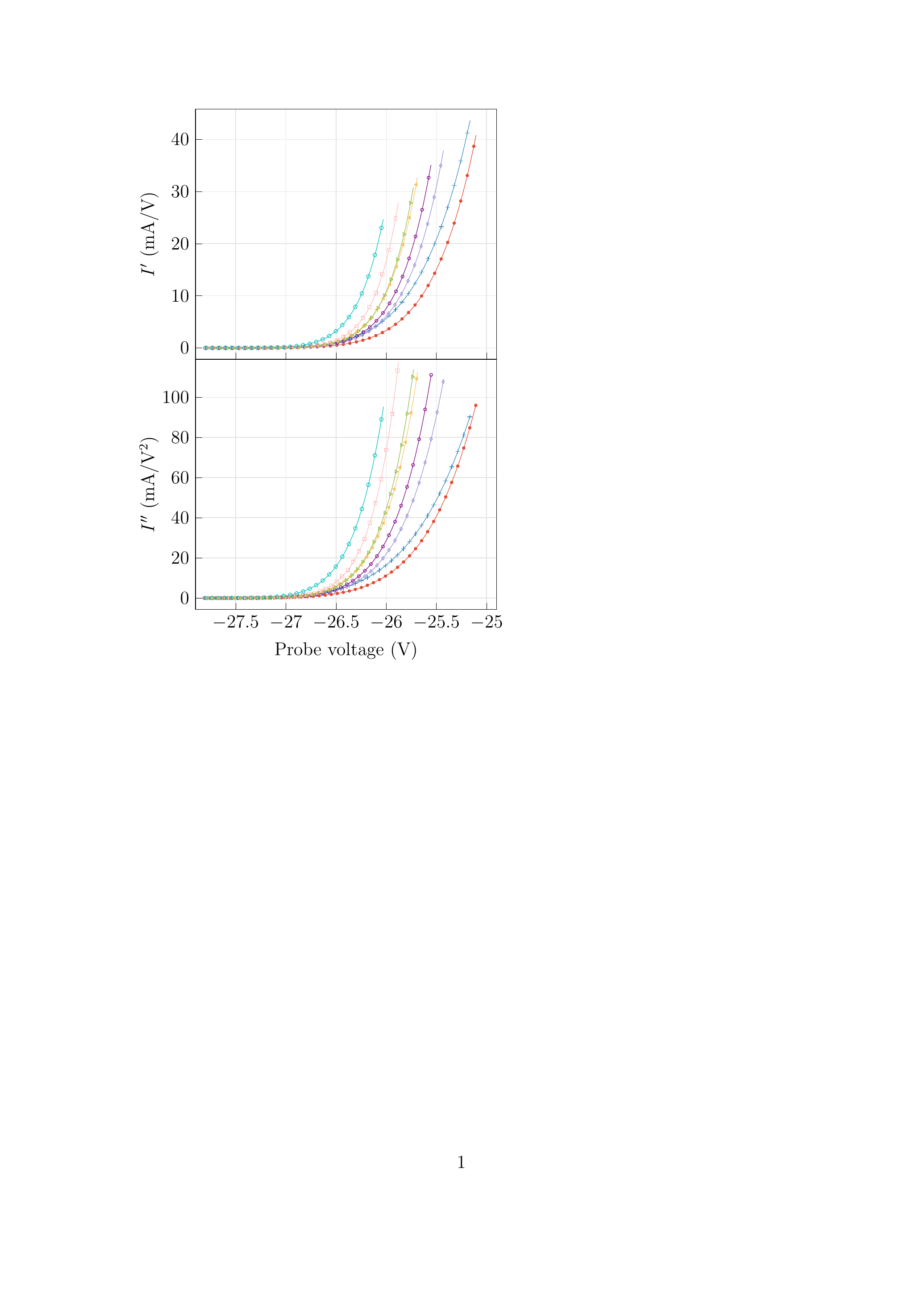}
\caption{ $I^\prime(V)$ and $I^{\prime \prime}(V)$ curves obtained from the analytical expression of the first and second derivatives of the function defined in Eq.~(\ref{ec:qiu}) with the parameters found in the corresponding fit, with the probe at different heights measured from the cathode: \textcolor{color6}{$\pentagon$} 9.4, \textcolor{color5}{\large$\square$} 8.2, \textcolor{color4}{$\triangleright$} 7.3, \textcolor{color3}{$\blacktriangle$} 6.2, \textcolor{violet}{\large$\circ$} 5.1, \textcolor{color2}{\small$\lozenge$} 4.3, \textcolor{color1}{\large$\plus$} 3.4 and \textcolor{color0}{\small$\bullet$} 2.2 mm. The symbols are introduced only to distinguish the different curves, which are continuous.}
\label{plotsIVqiu}
\end{figure}

\begin{figure}[H]
    \centering
    \includegraphics[width=0.65\linewidth]{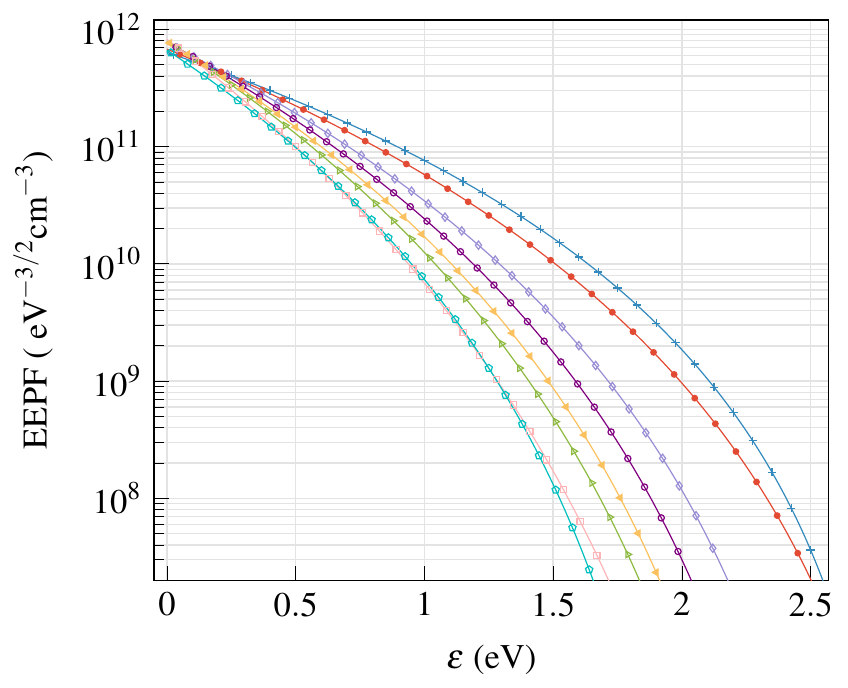}
    \caption{{\it{EEPF}}  curves obtained from the analytical expression of the second derivative using the function defined in Eq.~(\ref{ec:qiu}), evaluated at the corresponding fit parameters with the probe at different heights measured from the cathode: \textcolor{color6}{$\pentagon$} 9.4, \textcolor{color5}{\large$\square$} 8.2, \textcolor{color4}{$\triangleright$} 7.3, \textcolor{color3}{$\blacktriangle$} 6.2, \textcolor{violet}{\large$\circ$} 5.1, \textcolor{color2}{\small$\lozenge$} 4.3, \textcolor{color1}{\large$\plus$} 3.4 and \textcolor{color0}{\small$\bullet$} 2.2 mm. The symbols are introduced only to distinguish the different curves, which are continuous.}
    \label{plotsEEPqiu_logscale}
\end{figure}

The fitted curves using Eq.~(\ref{ec:qiu}) are shown in Fig.~\ref{fig-VI_Qiu} (using only two heights from the cathode). Figure~\ref{plotsIVqiu} presents the first and second derivatives of Eq.~(\ref{ec:qiu}) evaluated with the fitted parameters for each height, also the corresponding {\it{EEPF}} in semilog-scale is shown in Fig.~\ref{plotsEEPqiu_logscale}. The results are summarized in Table~\ref{tab:qqiu}. The errors from the table were estimated from the uncertainty of the nonlinear fitting and from the uncertainty in the $V_p$ value which comes from the SG filtering, both contributing approximately the same value. 

To obtain the {\it{EEPF}}, the analytic expression of $I^{\prime \prime}$ is used; this has the advantage, compared to the SG filter, of not having the numerical error given by the differentiation process. The expressions of $I^{\prime \prime}$ and the  {\it{EEPF}} (obtained from Eq.~(\ref{ec:Druyvesteynf})) are the following
\begin{equation}
    I^{\prime\prime}_Q(V)=\frac{AC_{q^*}}{4}B^2(3-q^*)(q^*+1)(1+B(q^*-1)(V-V_p))^{\frac{1}{q^*-1}-\frac{3}{2}}
\end{equation}
\begin{equation}
    f_Q(V)=\frac{4}{e^3S}\sqrt{\frac{m}{2}} I^{\prime\prime}_Q(V_p-V) \;, \, V>0
    \label{eq:fqiu}
\end{equation}

In conclusion, the use of Eq.~(\ref{ec:qiu}) to fit the experimental data has the advantage, compared to the SG filtering procedure, of determining the fitting parameters without ambiguities as a result of the nonlinear fitting, also the analytic expression of $I^{\prime \prime}$ can be used to calculate analytically the  {\it{EEPF}} and the other plasma parameters. However, this approach has the disadvantage that it is needed to previously calculate the plasma potential $V_p$, and also it dismisses the experimental $V>V_p$ values beforehand. The error previously estimated to $V_p$ is propagated, and affects greatly the final value of the plasma parameters. The SG filtering method (numerically differentiation), on the other hand, considers all the experimental data but, as we have seen in Sec. \ref{sec:SG}, it can be implemented by different options and this election is not obvious, therefore this method is somehow ambiguous to estimate plasma parameters. In the next section we consider another distribution in order to use the advantages of both methods, i.e. the use of an analytical expression to estimate the plasma parameters without ambiguities, and the use of all the experimental data to obtain smaller errors directly from the fit.

\begin{table}[H]
\caption{Summary table of the fit parameters found and the associated values of $n$ and $T$ calculated from Eqs. (\ref{ec:N}) and (\ref{ec:Te}) using the analytical expressions of the first and second derivatives of the expression defined in Eq.~(\ref{ec:qiu}) evaluated with the fitting parameters $q^*$, A and B. In this case the settings have a RMSE $\le 1\times 10^{-4}$. The $C_{q^*}$ parameter is not a free parameter of the fitting, it depends on $q^*$, $m_e$ and $m_i$, this parameter is added to the table only to ease the analysis. }
\label{tab:qqiu}
\centering
\begin{tabular}{ c c c c c c c }
\br
$h\!\pm\! 0.1$  & $q^*\!\!\!\pm\!0.02$ & $B\!\pm\! 0.1$  & $A\!\pm\!0.03$ & $C_{q^*}\!\pm\!0.9$  & $n\!\pm\!0.15$ & $T\!\pm\! 800$\\ 
(mm)  &  & (1/V) & (mA) &  & ($10^{11}\!/$cm$^3$) & (K)  \\ [0.5ex]
\mr
2.2  & 1.16 & 2.2 & 0.23 & 82.5 & 1.7 & 4800  \\
3.4  & 1.18 & 2.1 & 0.22 & 81.8  & 1.7 & 4600 \\
4.3  & 1.16 & 2.5 & 0.20 & 82.6  & 1.5 & 4400 \\ 
5.1  & 1.16 & 2.5 & 0.15 & 82.4  & 1.3 & 4000  \\ 
6.2  & 1.14 & 3.1 & 0.17 & 82.3  & 1.3 & 3900 \\ 
7.3  & 1.15 & 3.1 & 0.14 & 83.2  & 1.1 & 3700  \\ 
8.2  & 1.14 & 3.4 & 0.14 & 83.5  & 1.0 & 3700 \\
9.4  & 1.15 & 3.4 & 0.11 & 82.8   & 0.9 & 3400 \\
\br
\end{tabular}
\end{table}

\subsection{Using the $q$-Weibull distribution function: plasma potential as a parameter of the fit}\label{sec:qweibull}

\begin{figure}[htpb]
\centering
\includegraphics[width=0.4\linewidth]{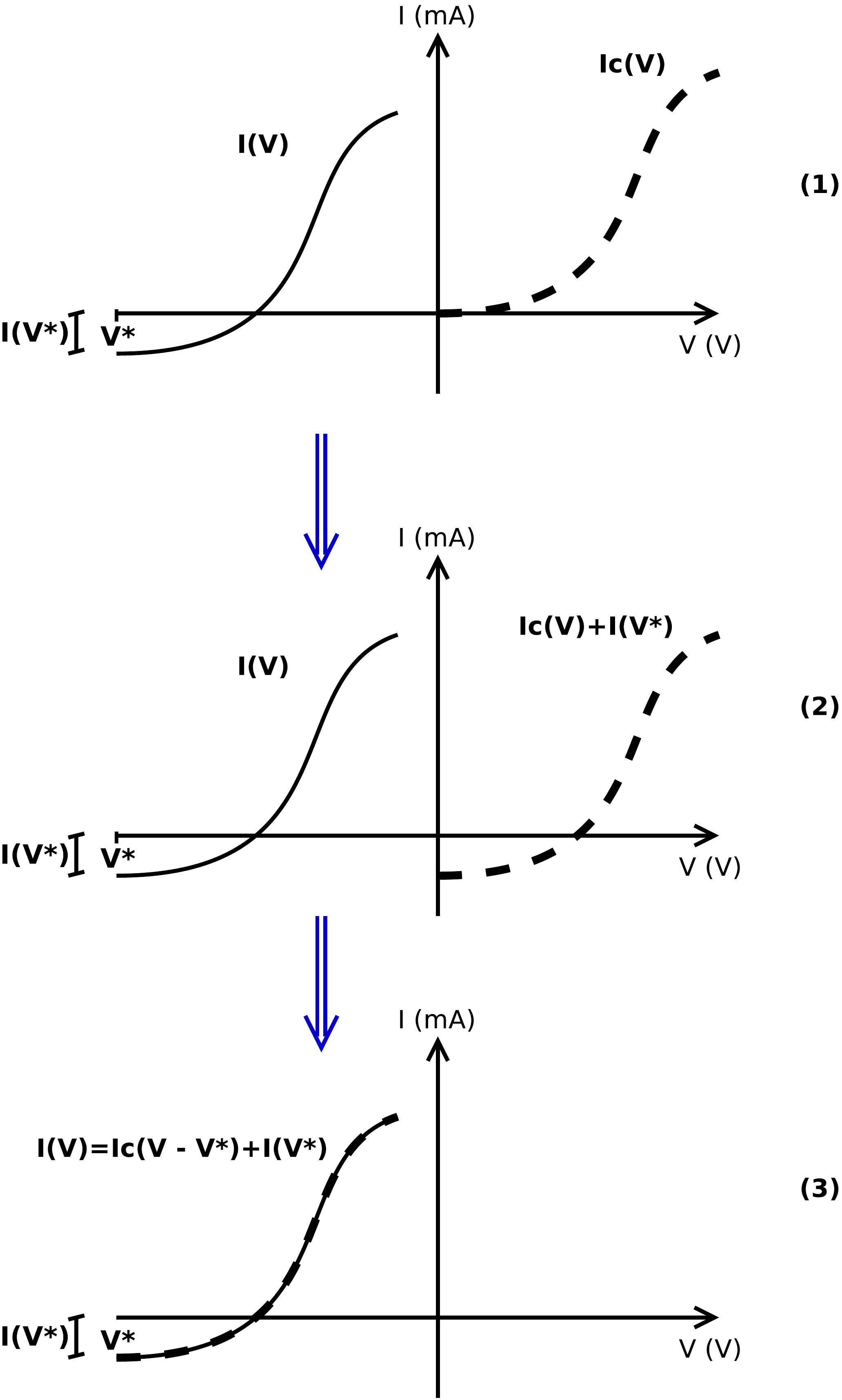}
\caption{Graphical description of the steps of the procedure in order to represent a Langmuir curve by using the $q$-Weibull cumulative function.}
\label{fig-QWprocedure}
\end{figure}

In this section, we consider some distribution functions with enough flexibility to represent the complete {\bf{I-V}} curves of this work. We propose a general procedure that can be used in other plasma conditions, and the aim of this approach is to determine the plasma parameters without ambiguities and also to extend the non-extensive $q$-exponential distribution which is not able to fit the complete {\bf{I-V}} curve. 
We first note that the experimental {\bf {I-V}} curves in Fig.~\ref{fig-iv} present a remarkable similarity to a cumulative distribution function (cdf) type. Then, we propose to use a cdf with enough flexibility to fit the {\bf {I-V}} curves, the restriction we must include is that the cdf has an inflection point (which corresponds to $V=V_p$). The $q$-exponential cdf does not satisfy this criterion, so we propose the $q$-Gaussian and $q$-Weibull distributions, which are natural generalizations of the $q$-exponential distribution\cite{PicoliJr2009}. The $q$-Weibull distribution includes an additional $r$ parameter, and in the cases $r=1$ and $r=2$ we recover the $q$-exponential and $q$-Gaussian distributions, respectively. The $q$-Weibull distribution also has been used in a variety of contexts in complex systems\cite{PicoliJr2009,Zhang2018,Abbas2020}, and it is closely linked to the Tsallis statistics\cite{Masi2005,Jizba2017,Lins2018,Frank2000}. We finally propose to fit the {\bf {I-V}} curves with the $q$-Weibull cdf, shown in Eq.~(\ref{ec:cdfQWorig}). It is important to note that this cdf is related to the {\it{EEPF}} through a second derivative, this is analyzed in Appendix C.

\begin{equation}
      I_c(V;V_p,q,r,A)=A\left(1-\left[1-(q-1) \left(\frac{V}{\lambda}\right)^r\right]^{\frac{q-2}{q-1}} \right)
      \label{ec:cdfQWorig}
\end{equation}
To implement Eq.~(\ref{ec:cdfQWorig}) we propose a slightly modified version which is shown in Eq.~(\ref{ec:cdfqWeibull}). To explain these modifications we note first that the $q$-Weibull cdf (Eq.~(\ref{ec:cdfQWorig})) is a cumulative function, this implies it starts from the origin of coordinates. Therefore, a shift from the origin of coordinates to the initial point of the $I-V$ curve must be made. This shift is shown in Fig.~\ref{fig-QWprocedure}. Subplot $(1)$ represents a complete {\bf {I-V}} Langmuir curve whose lower voltage is $V^*$ and the corresponding current is $I(V^*)$, it is also shown the theoretical $q$-Weibull cdf as a dashed curve ($I_c$). Subplot $(2)$ consists of a shift in the probe current axis of the $I_c$ curve, at a value of $I(V^*)$. This value will be a fitting parameter and will be called $I_s$, in the following we will suppose that $I_s$ does not depend on $V$, this is a rough approximation that will allow us to neglect this contribution in the derivatives of the function. In the case $V^*<<V_f$, the $I_s$ parameter can represent the ion saturation current. Step $(3)$ consists of a shift in the voltage axis of the $I_c$ curve, at a value of $V^*$. After step $(3)$, the corresponding curve can represent the complete {\bf{I-V}} Langmuir curve. Finally, we replace the $\lambda$ parameter to parameters: $V_p$, $q$, and $r$. This is achieved by obtaining the second derivative of the cdf from step $(3)$ and matching to zero, in this value it is verified $V=V_p$, then we isolate the variable $\lambda=\lambda(V_p,q,r)$. Equation (\ref{ec:cdfqWeibull}) summarizes the procedure for the case $r=1$, but the general case including $r=1$ is analyzed in Appendix B.

In summary, Eq. (\ref{ec:cdfqWeibull}) includes $V_p$, $q$, $r$, $A$ and $I_s$ as fitting parameters. The initial values of these parameters to find the numerical convergence, as we explain in the following, can be easily found.

\begin{equation}
  I_{QW}(V;V_p,q,r,A,I_s)= A\left(1-\left[1-\frac{(q-1)(r-1)}{(q-1)(r-1)-r} \left(\frac{V-V^*}{V_p-V^*}\right)^r\right]^{\frac{q-2}{q-1}} \right)+I_s
  \label{ec:cdfqWeibull}
\end{equation}

The initial values of the parameters must be carefully selected to ensure the convergence of the nonlinear fit\cite{Xu2017}. The initial values of $V_p$ and $A$ can be obtained from the visualization of the {\bf{I-V}} curve, the $V_p$ value can be estimated from the convexity change, and the $A$ value from the subtraction of the electron saturation current with the ion saturation current, i.e. $A\approx (I(V\gg V_p) - I(V\ll V_f))$. On the other side, the initial values of $q$ and $r$ can be selected in the ranges of $(0.5,1.5)$ and $(1,10)$, respectively. The $I_s$ value can be initially chosen as $I(V^*)$. Taking into account these criteria, we can say that convergence is easily reached. The only error that may appear after convergence is reached consists of the error itself from the fit, because the analytic expression of the second derivative can be used, and the numerical differentiation is not needed. Moreover, an advantage of this procedure is that the $V_p$ value  can be determined without ambiguities as a result of the fit.
The analytic expression of the first and second derivatives, with $r\neq 1$, are shown in Eqs.~(\ref{ec:pdfqWeibull}) and (\ref{ec:derpdfqWeibull}), respectively.

\begin{equation}
  I^\prime_{QW}\!(V\!;V_p,q,r,A)\!=\frac{A(q-2)(r-1)r}{(q-1)(r-1)-r} \frac{(V-V^*)^{r-1}}{(V_p-V^*)^r} \!\left[1\!-\!\frac{(q-1)(r-1)}{(q-1)(r-1)-r}\! \left(\frac{V-V^*}{Vp-V^*}\right)^r\right]^{\frac{1}{1-q}}   \; ,
  \label{ec:pdfqWeibull}
\end{equation}

\begin{eqnarray}
  I^{\prime\prime}_{QW}\!(V\!;V_p,q,r,A)\!=  & A \left[1-\frac{(q-1)(r-1)}{(q-1)(r-1)-r}\left( \frac{V-V^*}{V_p-V^*}\right)^r  \right]^{\frac{q}{1-q}} \cdot \nonumber \\
    & \frac{r(r-1)^2(q-2)}{(q-1)(r-1)-r}\frac{(V-V^*)^{r-2}}{(V_p-V^*)^r} \left(1- \left(\frac{V-V^*}{V_p-V^*}\right)^r\right)
  \label{ec:derpdfqWeibull}
\end{eqnarray}

Also, the {\it{EEPF}} can be obtained by the variable change $V$ to $V_p-V$, as is required by Druyvesteyn in Eq.~(\ref{ec:Druyvesteynf}), obtaining, to $r\neq 1$ and $V>0$, Eq.~(\ref{ec:fqWeibull}).

\begin{eqnarray}
  \label{ec:fqWeibull}
  \fl  f_{QW}\!(V\!;V_p,q,r,A)\!= &
     \frac{4A}{e^3S}\sqrt{\frac{m}{2}}\left[1-\frac{(q-1)(r-1)}{(q-1)(r-1)-r}\left( \frac{V_p-V^*-V}{V_p-V^*}\right)^r  \right]^{\frac{q}{1-q}} \cdot \nonumber \\
     & \frac{r(r-1)^2(q-2)}{(q-1)(r-1)-r}\frac{(V_p-V^*-V)^{r-2}}{(V_p-V^*)^r} \left(1- \left(\frac{V_p-V^*-V}{V_p-V^*}\right)^r\right) \; .
\end{eqnarray}

\begin{figure}[H]
\centering
\includegraphics[width=0.65\linewidth]{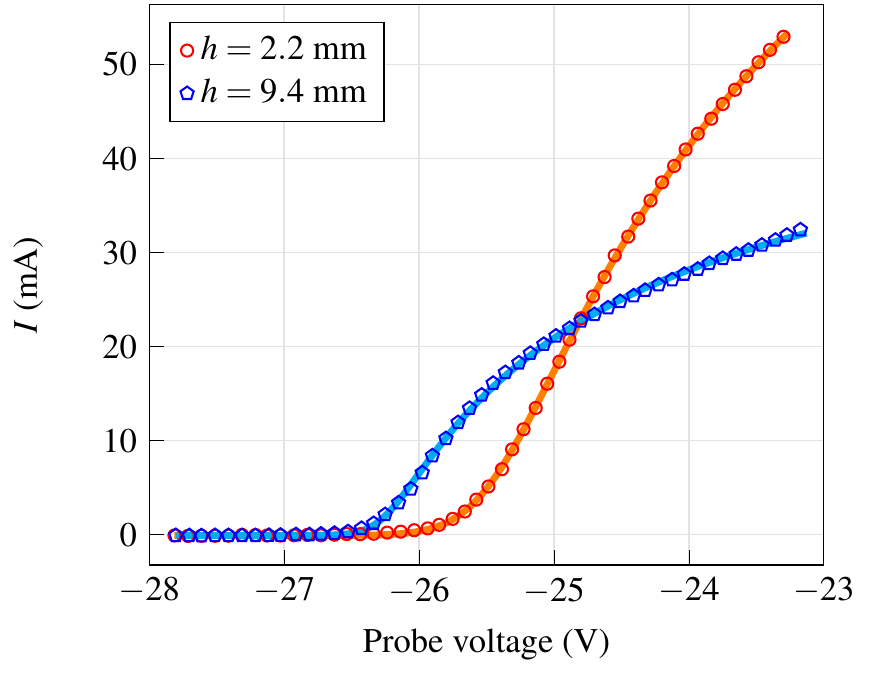}
\caption{(Symbols) Experimental {\bf {I-V}} measurements, for the probe at two different distances from the cathode. (Continuous lines) Fit of the experimental {\bf {I-V}} data using the $q$-Weibull cumulative distribution.}
\label{IV-qWeibull}
\end{figure}

\begin{figure}[H]
    \centering
    \includegraphics[width=0.85\linewidth,trim= 2.5cm 14.5cm 8.5cm 1.8cm]{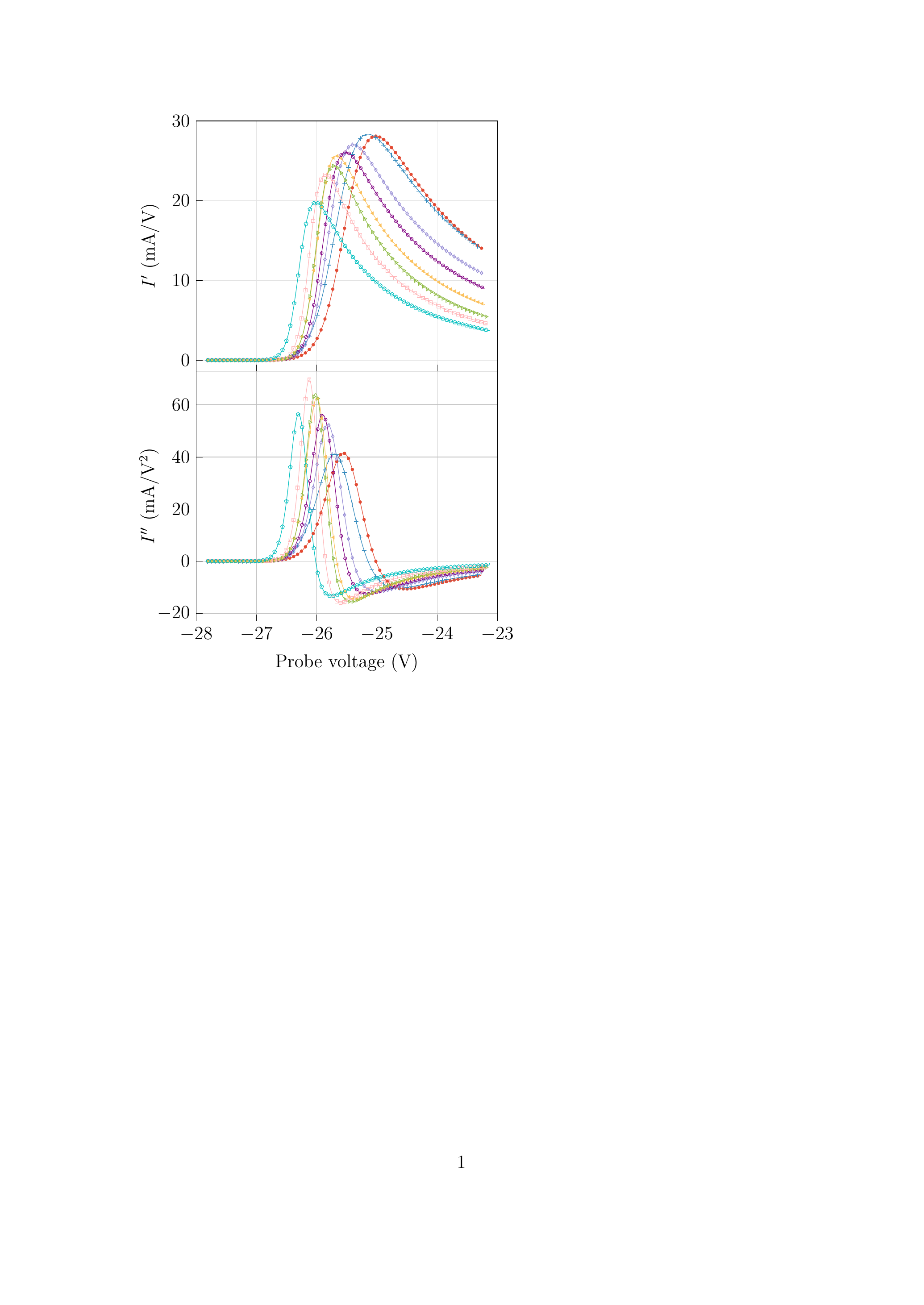}
    \caption{$I^\prime(V)$ and $I^{\prime \prime}(V)$ curves obtained from the analytical expression of the first and second derivatives of the {\it{q}}-Weibull cumulative distribution and the parameters found in the corresponding fit, with the probe at different heights: \textcolor{color6}{$\pentagon$} 9.4, \textcolor{color5}{\large$\square$} 8.2, \textcolor{color4}{$\triangleright$} 7.3, \textcolor{color3}{$\blacktriangle$} 6.2, \textcolor{violet}{\large$\circ$} 5.1, \textcolor{color2}{\small$\lozenge$} 4.3, \textcolor{color1}{\large$\plus$} 3.4 and \textcolor{color0}{\small$\bullet$} 2.2 mm,  measured from the cathode. The symbols are introduced only to distinguish the different curves, which are continuous.}
    \label{fig:ivAbbas}
\end{figure}

\begin{figure}[H]
    \centering
    \includegraphics[width=0.8\linewidth]{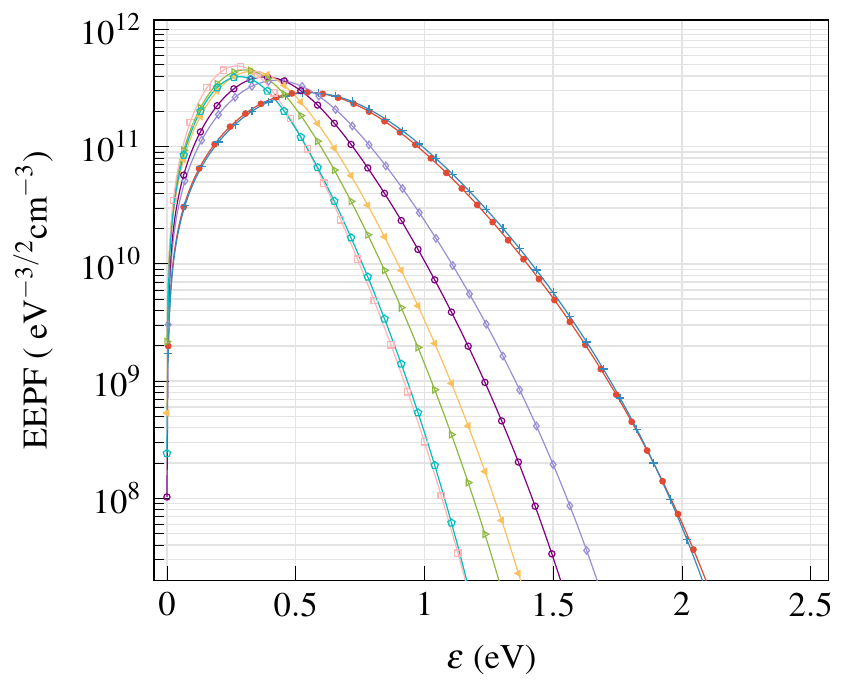}
    \caption{{\it{EEPF}}  curves obtained from the analytical expression of the second derivative of the {\it{q}}-Weibull cumulative distribution, evaluated at the corresponding fit parameters with the probe at different heights measured from the cathode: \textcolor{color6}{$\pentagon$} 9.4, \textcolor{color5}{\large$\square$} 8.2, \textcolor{color4}{$\triangleright$} 7.3, \textcolor{color3}{$\blacktriangle$} 6.2, \textcolor{violet}{\large$\circ$} 5.1, \textcolor{color2}{\small$\lozenge$} 4.3, \textcolor{color1}{\large$\plus$} 3.4 and \textcolor{color0}{\small$\bullet$} 2.2 mm.  The symbols are introduced only to distinguish the different curves, which are continuous.}
    \label{fig:eepfabbas}
\end{figure}

The fitted curves using Eq.~(\ref{ec:cdfqWeibull}) are shown in Fig.~\ref{IV-qWeibull}. Figure~\ref{fig:ivAbbas} presents the first and second derivative of the {\bf{I-V}}  data evaluated with the fitted parameters for each height, also the corresponding {\it{EEPF}}, obtained from the analytic expression of the second derivative, is shown in Fig.~\ref{fig:eepfabbas}. We dedicate a comment for the depleted shape observed at low energies ($\Delta$). With the fit procedure, we have eliminated the problem of the convolution effect, but the effects of electron reflection, secondary-electron emission, inhomogeneity of the probe work function along the probe collecting surface are still present. We observe in Fig.~\ref{fig:eepfabbas} that the values of $\Delta \approx (0.3-0.6)$ eV are similar to those obtained in Sec.\ref{sec:SG}, this means that the convolution effect is not responsible for the actual value of $\Delta$ in our measurements.

The results are summarized in Table \ref{tab:qweibull}. The plasma parameters $n$ and $T$ are obtained from the analytic expression of the {\it{EEPF}} by using Eqs.~(\ref{ec:N}) and (\ref{ec:Te}). 
The errors corresponding to $n$ and $T$ were estimated from the uncertainties of the fitting parameters. To be more precise, we have calculated the variations in the values of $n$ and $T$ when the fitting parameters are moved in the range of its uncertainties, then we selected a bound for the uncertainties of $n$ and $T$. In the table, we also show the values of $V_f$, which are obtained from Eq.~(\ref{ec:cdfqWeibull}) solving for the potential with $I_{QW}=0$.

From Table \ref{tab:qweibull} we observe that $q$-Weibull distribution has a smaller root-mean-square error (RMSE) compared with trimmed data of Sec.~\ref{sec:qiu}, which implies better results in predicted plasma parameters. 

\begin{table}[H]
\caption{Summary table of the fit parameters of the $q$-Weibull distribution and the associated values of $T$ and $n$ calculated from Eqs.~(\ref{ec:N}) and (\ref{ec:Te}) by using the analytic expression of the second derivative of the probe current and the fitted parameters. In this case, the curve fitting has RMSE $\le 2\times 10^{-5}$. }
\label{tab:qweibull}
\centering
\begin{tabular}{ c c c c c c c c c}
\br
$h\!\pm\! 0.1$ & $V_{p}\!\pm\! 0.005$ & $V_f\!\pm\! 0.01$ & $q\!\pm\! 0.003$  & $r\!\pm\! 0.07$ & $A\!\pm\! 0.003$ & $I_S\!\pm\! 0.01$  & $n\!\pm\!0.05$ & $T\!\pm\! 150$ \\ 
(mm) & (V) & (V) & &  & (A) & ($mA$) & ($10^{11}\!/$cm$^3$) & (K) \\ [0.5ex]
\mr
2.2 & -25.023 & -26.38 & 1.936  & 11.23 & 0.138 & -0.28 & 1.48 & 5100 \\
3.4 & -25.153 & -26.49 & 1.942  & 10.37 & 0.157 & -0.33 & 1.50  & 5200 \\
4.3 & -25.400 & -26.59 & 1.956  & 13.64 & 0.130 & -0.26 & 1.25 & 3900 \\
5.1 & -25.523 & -26.72  & 1.953  & 14.41 & 0.107 & -0.25 & 1.14 & 3500 \\
6.2 & -25.671 & -26.61 &1.947  & 15.42 & 0.081 & -0.25 & 1.05 & 3100 \\
7.3 & -25.720 & -26.62 & 1.940  & 15.42 & 0.063 & -0.24 & 0.96 & 2800 \\
8.2 & -25.856 & -26.68 & 1.944  & 17.59 & 0.057 & -0.1 & 0.86 & 2500 \\
9.4 & -26.020 & -26.82 & 1.944  & 14.89 & 0.052 & -0.09 & 0.75 & 2700 \\

\br
\end{tabular}

\end{table}

\section{Discussion}\label{sec:discussion}

We have analyzed the {\bf{I-V}} data using the Druyvesteyn equation to obtain the electron density and temperature with three different methodologies: the first one with a numerical differentiation; the second one with a known expression from recent works based on the $q$-exponential distribution; and the last one with the $q$-Weibull cumulative distribution function.
On the one hand, with the first methodology, the numerical filters in general made use of the complete  {\bf {I-V}} curve to numerically obtain the second derivative and only in the calculation of the {\it EEPF}, the data for which $V>V_p$ are trimmed. Many methods can be used to obtain the second derivative, we decided to study the Savitzky-Golay filters (Sec.~\ref{sec:SG}). We have shown that this filtering procedure can be carried out by different filter parameters, which produce different results. Therefore, the estimated plasma parameters obtained with this methodology are somehow ambiguous. 
On the other hand, with the second and third methodologies, we have made nonlinear fittings by using analytic expressions I(V), which depend on certain fitting parameters. Then, we have used the analytical expressions of their second derivatives to calculate the electron temperature and density. This procedure allows us to obtain without ambiguities the plasma parameters as a result of the fit.
In the second methodology, we use the $q$-exponential distribution based on recent works (Sec.~\ref{sec:qiu}). We notice that it is needed to eliminate some experimental data in the {\bf {I-V}} curve to find numerical convergence, specifically the probe voltages $V>V_p$ need to be trimmed beforehand. One of the advantages of this second methodology, as it was mentioned in the previous paragraph, is that the plasma parameters are determined without ambiguities as a result of the fit, however, the main disadvantage is that the $V_p$ value needs to be calculated or estimated before than the fitting procedure is carried out, because with the complete {\bf {I-V}} curve, the convergence is not reached. This has as consequence that an additional error is added to the plasma parameters. We have shown that the second methodology yields greater errors in the plasma parameters than the Savitzky Golay procedure, this is shown in Tables \ref{tab:druv} and \ref{tab:qqiu}.
The third methodology consists of using the $q$-Weibull distribution function (Sec.~\ref{sec:qweibull}). This function is empirically introduced as the simplest generalization of the $q$-exponential distribution which can fit the complete \textbf{I-V} Langmuir curve measured.
With this approach, we obtain the plasma parameters without ambiguities as a result of the fit, and we also use the complete {\bf {I-V}} Langmuir curve. The numerical filtering procedure is not needed to obtain $V_p$, $q$, $r$, $A$ and $I_s$, because these parameters are directly obtained as a result of the fit. This allows to get (from the analytical expression of the second derivative by using expressions from Sec.~\ref{sec:druyvesteyn}) the $n$ and $T$ plasma parameters with greater precision, i.e. with smaller errors, compared to the first two methodologies, as is shown in Table \ref{tab:qweibull}. It is important to mention that this function can have limitations to represent the ion current contribution, however, the $q$-Weibull allows us to obtain uniquely and with great precision the plasma parameters, as it is discussed in some detail below in this section. In Appendix D, we discuss the variation of the plasma parameters with the initial voltage $V^*$, we observe that plasma parameters remain roughly constant with $V^*$, and this means that the $q$-Weibull has an almost horizontal behavior in the range from $V^*$ to $V_f$. This result may indicate that the $q$-Weibull function can only fit the electron contribution of the \textbf{I-V} curve.

\begin{figure}[H]
\begin{subfigure}{.5\textwidth}
  \centering
\includegraphics[width=0.9\linewidth]{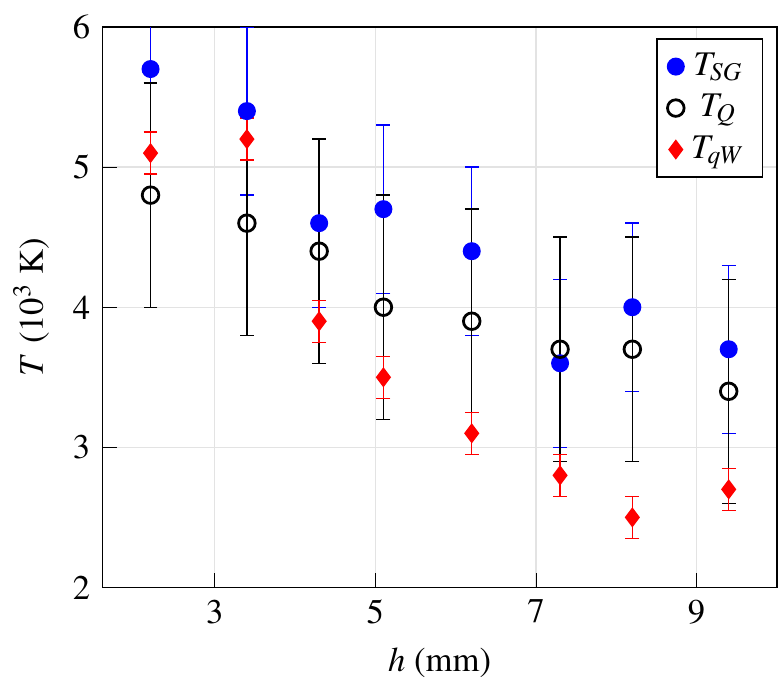}
\caption{Temperature.}
\label{fig:temperature}
\end{subfigure}
\begin{subfigure}{.5\textwidth}
  \centering
\includegraphics[width=0.94\linewidth]{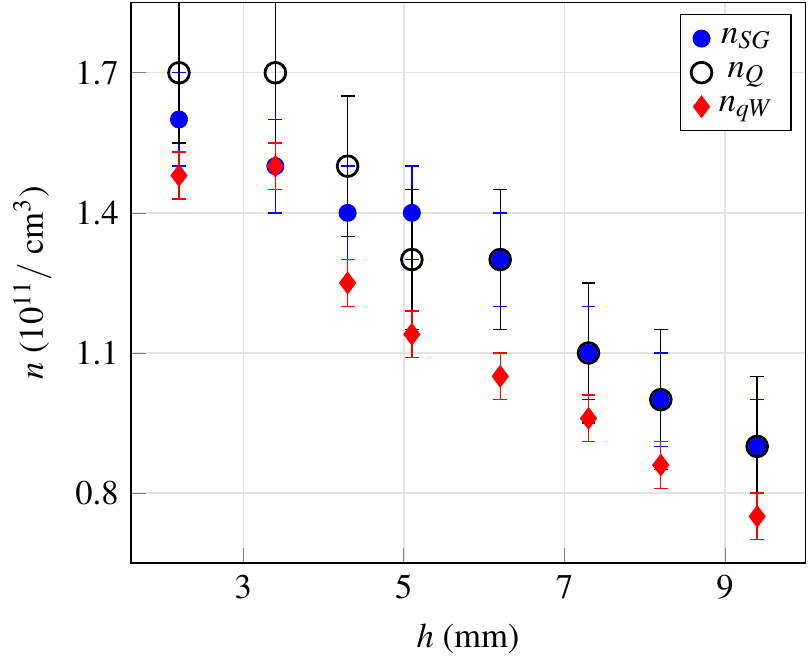}
\caption{Density.}
\label{fig:density}
\end{subfigure}
\caption{Comparison of temperatures and densities obtained by means of traditional numerical derivation ($SG$) from Sec.~\ref{sec:druyvesteyn}, the non-extensive distribution described in Sec.~\ref{sec:qiu} ($Q$) and the $q$-Weibull ($qW$) cdf from Sec.~\ref{sec:qweibull}.}
\label{fig:tempdensity}
\end{figure}

In the following, we compare the results from the three methodologies. Figure (\ref{fig:tempdensity}) shows the comparison of temperatures and densities with the three proposals described in this work. Section \ref{sec:druyvesteyn} shows that plasma parameters $T$ and $n$ are very sensitive to the variation of $V_p$, this means that an estimation of the exact value of $V_p$ is key. With the first methodology (SG filtering), the $V_p$ value changes according to the filter parameters and the chosen filter. This implies that the precise value of $V_p$ is not known, and it contributes almost entirely to the errors shown in Table \ref{tab:druv}.
Otherwise, with the $q$-Weibull methodology, the $V_p$ value is determined without ambiguities as a result of the fit. Also, the error in $V_p$ is lower compared to the first methodology, this is translated into smaller errors for $T$ and $n$. 

\begin{figure}[htpb]
\centering
\includegraphics[width=0.7\linewidth]{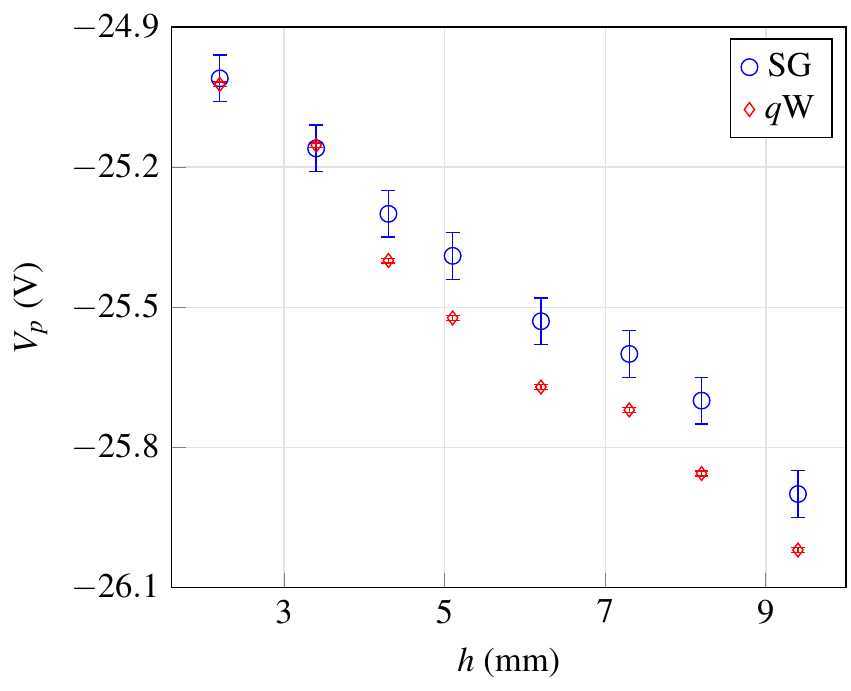}
\caption{Comparison of $V_p$ values with the numerical filtering with SG and the $q$-Weibull approach.}
\label{fig:vpvalues}
\end{figure}

To evaluate the importance of the $V_p$ value, in Fig. \ref{fig:vpvalues} we show the $V_p$ values obtained by using the numerical differentiation with the SG filter and the fit with the $q$-Weibull function. We observe that for the heights greater than 4 mm, the $q$-Weibull approach underestimates the $V_p$ value compared with the numerical differentiation using the SG filters. We also observe that at these heights, the temperature and density obtained from $q$-Weibull approach are lower compared with the achieved with the numerical differentiation using the SG filers. We associate this underestimation of $T$ and $n$ to the underestimation of $V_p$. At this stage, we are able to conclude that the $V_p$, $T$ and $n$ values obtained from the $q$-Weibull approach are more precise, i.e. lower uncertainties, compared with the numerical differentiation and the $q$-exponential approach.

\begin{figure}[htpb]
\centering
\includegraphics[width=0.7\linewidth]{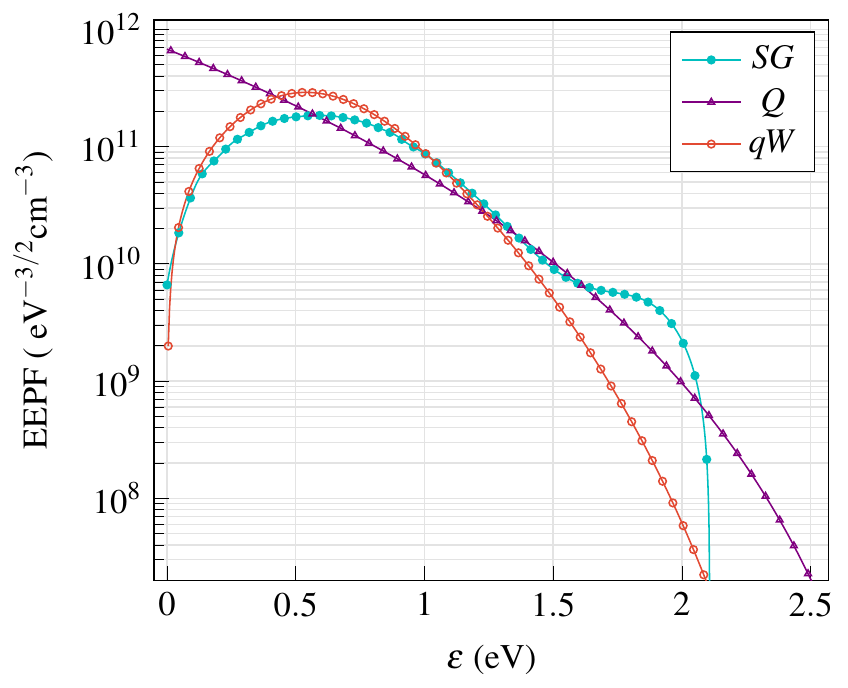}
\caption{Comparison of the  {\it EEPF}s  after applying the three different methodologies at 2.2 mm height.}
\label{fig:eepf22}
\end{figure}

Figure \ref{fig:eepf22} shows the  {\it{EEPFs}}  of the three methodologies implemented in this work, considering only the 2.2 mm height. We can observe that the Qiu {\it{EEPF}}  has a maximum at zero energies, this can be an advantage if a maximum at zero energies is expected. 
Mathematically, this is due to the fact that Eq.~(\ref{eq:fqiu}) is based on the $q$-exponential distribution, which does not present a concavity change. However, in this work the \textbf{I-V} curve is mantained without corrections and the depleted shape at low electron energies is present. The $q$-Weibull {\it{EEPF}} can represent this low energy zone consistently with the SG method. The SG {\it{EEPF}} shows a non-physical rounded shape around 2eV, this shape changes with the chosen filter parameters, therefore it depicts the presence of numerical errors.

To finish this section, we are going to discuss the limitations of the $q$-Weibull methodology. In this work, we have considered the experimental values of a Langmuir cylindrical probe from a few Volts below the floating potential $V_f$ up to a few Volts above the plasma potential $V_p$. The high flexibility of the $q$-Weibull function\cite{Xu2017} allows it to represent a wide range of probe voltages, in particular the curves we analyzed in this work, which have a non-Maxwellian distribution, can be depicted with sufficient accuracy by the $q$-Weibull distribution. This means that all the physics concerning the calculation of the plasma parameters can be described, collectively, with the $q$-Weibull function. Specifically, we have taken measurements with the probe located at different heights near the cathode with certain plasma conditions mentioned in Sec. \ref{sec:ivcurve}, where we have found that the plasma is in the conventional thin sheath subregime. This regime has shown to be well represented by the $q$-Weibull distribution. However, we have not yet tested the use of this function with other plasma conditions or a wider range of voltages. Probably, the $q$-Weibull function may fail to represent a more extensive range of voltages, i.e. high ion energies ($V\ll V_f$), which correspond with the ion current contribution. In this respect, the q-Weibull function neglects the ion contribution and the approximation $d^{2}I_{e,ret}/dV^{2} =I_{e,ret}^{\prime\prime}$ mentioned in Sec. \ref{sec:druyvesteyn} is taken implicitly. In this direction, the $q$-Weibull function deserves further research regarding the ion current contribution that is beyond this work. Furthermore, the $q$-Weibull function may have problems to represent peaks in the {\it{EEPF}}, like those coming from electron beams\cite{Li2020} or primary electron spikes \cite{Longhurst1981}. In the case of Nitrogen plasma, the $q$-Weibull function may fail to represent dips and peaks in the EEPF at higher electron energies (above $5$ eV), as it was reported in Refs.\cite{Winkler1973ZurMBI,Winkler1977ZurMBII,Pfau1977ZurMBIII}, however these effects are of some order of magnitude below the maximum value of the EEPF, and therefore could contribute only a small amount compared with the low-energy electron peak which is the part of the curve which better fits the $q$-Weibull function. This may be a subject of future investigations.

With the $q$-Weibull methodology, the system is analyzed by one ``mean'' temperature obtained by integrating the second derivative of the probe characteristic, as opposed to other approaches that uses two temperatures, for example in a Bi-Maxwellian distribution founded in a low-pressure RF discharge with noble gases where the Ramsauer effect is present\cite{Godyak1992RF}. In summary, the $q$-Weibull may have problems to represent some effects like peaks in the  {\it{EEPF}} or more complex shapes, such as the ones mentioned before, but it has enough flexibility to fit to a wide class of functions. We can conclude that, despite these limitations, if only the ``mean'' plasma parameters are needed, the $q$-Weibull can be a useful alternative to determinate plasma parameters due to its high precision. Otherwise, if the particular study of these peaks or complex {\it{EEPF}} shapes are needed, it is advisable to reside in other method, such as numerical differentiation. These peaks and complex shapes have been studied based on microscopic models \cite{Hippler2008}.

In spite of the limitations mentioned, we consider that the $q$-Weibull distribution has enough flexibility to be used in different contexts, as is shown in many applications that have similar {\it{EEPF}}s compared to the ones we present in this work, for example in plasma space \cite{Olson2010,Abe2013,Samaniego2019}, plasma fusion \cite{Nie2018}, dusty plasmas \cite{Denysenko2015} and cold plasma physics \cite{Kortshagen1994,Maresca2002,Denysenko2004,Takahashi2011,Boswell2015,Darian2019} in general. We expect that the $q$ and $r$ parameters can be useful to determine the plasma operating regimes and the discharge zones, and to measure the degree of departure of the system from the Maxwellian distribution. Particularly in this work, we have found that all the measurements at different heights near the cathode have a $q$ value that is in the range of $(1.943\pm 0.006)$, this corresponds to a 0.3 $\%$ of relative error, which is small enough to reach the conclusion that all heights share the same $q$ parameter. There is still work to be done in order to consider the $q$-Weibull methodology both in experimental and theoretical approaches to clarify
the physical meaning of the $q$ and $r$ parameters. From the experimental approach, a further analysis can be made to obtain the $q$ and $r$ parameters with different plasma conditions and discharge zones, in particular the region of the ion current contribution ($V\ll V_f$) needs to be studied in detail to determinate if the $q$-Weibull can represent the ion contribution. From the theoretical approach, more efforts can be made to link the non-extensive statistics with the Langmuir probe measurements.

Finally, these considerations could be an important step to provide a more precise physical interpretation of the $q$ and $r$ parameters in the context of non-extensive statistics, and also to include these parameters with the usual $V_p$, $T$, $n$ and $V_f$ plasma parameters.

\section{Conclusions}\label{sec:conclusion}

In this work, we have presented the possibility of identifying the {\bf{I-V}} Langmuir curve with a function of cumulative type. For the first time, we introduce the $q$-Weibull cumulative function to fit a wide range of voltages from the {\bf{I-V}} Langmuir curve, from a few Volts below the floating potential ($V_f$) up to a few Volts above the plasma potential ($V_p$). This function is motivated by recent works in nonextensive statistics, which use a $q$-exponential distribution to represent non-Maxwellian {\it{EEPF}}s. 

We applied the $q$-Weibull procedure to fit the {\bf{I-V}} curves measured at different distances from the cathode on a N$_2$ cold plasma.

By using this procedure, the convergence of the fit is easily achieved by selecting the initial parameters from the observation of the \textbf{I-V} curve. Based on the analysis of these measurements, we demonstrated that the approach using the $q$-Weibull function has many advantages compared with the usual numerical differentiating method. In particular, we compared different variants of numerical differentiation by using the Savitzky Golay filters with the $q$-Weibull approach. On the one hand, we found that the plasma parameters $V_p$, $T$ and $n$ were very sensitive to the chosen filter parameters of the SG filters. This allowed us to show that the plasma parameters $V_p$, $T$ and $n$ are not uniquely determined when the numerical method was used. On the other hand, with the $q$-Weibull approach, the plasma potential $V_p$ was obtained directly from the fit, and the plasma parameters $T$ and $n$ were uniquely determined by evaluating their analytic expressions using the parameters $V_p$, $q$, $r$ and $A$ previously found in the fit. Therefore, we fit directly the {\bf{I-V}} curve and the numerical differentiations were not needed, as a consequence of this approach, the unique error came from the non-linear fit of the $q$-Weibull function, finally this yielded to smaller uncertainties of the plasma parameters $V_p$, $T$ and $n$ compared with the usual numerical differentiation using SG filters.

The $q$-Weibull function has five parameters $V_p$, $q$, $r$, $A$, and $I_s$. We proposed, in particular, that the $q$ and $r$ parameters could be related to the degree of nonextensivity of the system, the last one composed by the plasma and probe measurement, in other words, the $q$ and $r$ parameters could be related to the degree of departure of the system from the Maxwellian distribution. However, the physical interpretation of these nonextensive parameters requires future investigation. 
Finally, we can conclude that in case that only one unique value for each plasma parameter of the system is required, this method based on the fit with the $q$-Weibull cumulative distribution has proved to be very precise.

\section*{Acknowledgements}
This work was partially supported by Argentina National Scientific and Technical Research Council (Consejo Nacional de Investigaciones Cient\'ificas y T\'ecnicas, CONICET). The authors are thankful to Dr. Lucio Isola, Hern\'an Rindizbacher, Ing. Aldo Marenzana and Javier Cruce\~no for their essential collaboration in the experimental set-up. We thank Michael O. Thompson of
Cornell University for their help and support with the use of Genplot.  
We are grateful to our linguistic reviewer, Sabrina María Cid Sacramone, for her suggestions on the style and grammatical correctness of our English. We are also grateful to the anonymous referees, who through enriching discussions helped us to improve this work.

\section*{Appendix A: 3D \textit{EEPF} from Tsallis statistics }
 
The Tsallis entropy\cite{Tsallis1988} is
\begin{equation}
  S_q[p]= \frac{k}{q-1}\left( 1- \sum_{i=1}^N p_i^q \right)  ,
  \label{ec:Tsallisentropy}
\end{equation}
where $\{ p_i \}$ are the probabilities of the microscopic configurations, $k$ is the Boltzmann constant, $N$ is the total number of possible (microscopic) configurations, and $q$ is a real-valued parameter.

Consider the conservation conditions, which can be referred to as generalized spectrum and generalized internal energy respectively\cite{Tsallis1988}:
\begin{equation}
  \sum_{i=1}^N p_i = 1\ \ \ ,\ \ \  \sum_{i=1}^N  p_i  \varepsilon_i= U_q \ \ \ \  \text{where}\ \  \varepsilon_i, U_q \in R \,.
\end{equation}

The extremization of $S_q$ with these conditions yields to the $q$-exponential distribution

\begin{equation}
    p_i=\frac{1}{Z_q}\left[1-(q-1)\frac{\varepsilon_i}{kT}\right]^{\frac{1}{q-1}},
\end{equation}
with
\begin{equation}
    Z_q\equiv \sum_{i=1}^N \left[1-(q-1)\frac{\varepsilon_i}{kT}\right]^{\frac{1}{q-1}} \,.
\end{equation}

Using this non-extensivity context applied to plasmas, Silva\cite{Silva1998} proposed the 3D electron energy distribution:

\begin{equation}
  f_{3D_{Silva}}(\varepsilon)=B_q\left(1-(q-1)\frac{\varepsilon}{kT}\right)^{\frac{1}{q-1}}
\label{ec:3DSilva}
\end{equation}

where $B_q$ is a normalization constant\cite{Silva1998}.

Besides, Qiu and Liu\cite{Qiu2018} in 2018 propose the 3D electron energy distribution:
\begin{equation}
  f_{3D_{Qiu}}(\varepsilon)=\left(\frac{m}{2\pi kT}\right)^\frac{3}{2} A_q \left(1-(q-1)\frac{\varepsilon}{kT}\right)^{\frac{2-q}{q-1}} \, ,
  \label{ec:3DQiu}
\end{equation}


where $A_q$ is a normalization constant \cite{Qiu2018}.

\section*{Appendix B: General expression for the current probe by using the q-Weibull distribution, splitting the case $r=1$}

Following only steps $(1)$, $(2)$ and $(3)$ from the procedure explained in Sec. \ref{sec:qweibull}, we can express the probe current in a general way   

\begin{equation}
  I_{QW}(V;V_p,q,r,A,I_s)=A\left(1-\left[1-(q-1) \left(\frac{V-V^*}{\lambda}\right)^{r}\right]^{\frac{q-2}{q-1}} \right)+ I_s \; .
  \label{ec:cdfqWeibulllambda}
\end{equation}

From step $(4)$, we can express $\lambda$ as a function of $V_p$, $q$, and $r$, including also the case $r=1$ by
\begin{equation}
  \lambda= \begin{cases}
  (V_p-V^*) \left(\frac{(q-1)(r-1)-r}{r-1}\right)^{\frac{1}{r}} & ; r\neq 1  ,  \forall q\\
  (V_p-V^*) (q-1) & ; r = 1  ,  q\neq 1\\
  \lambda & ; r=1 ,  q =1
  \end{cases}
  \label{ec:lambda}
\end{equation}

A comment can be made to the case $r\neq 1$, $\forall q$, when considering $q=1$, it needs to be solved by limits and using the l'Hôpital rule. Other observation is that in the case $q=1$ with $r=1$, the $\lambda$ parameter cannot be expressed as a function of $V_p$ because Eq.~(\ref{ec:cdfqWeibulllambda}) has not an inflection point to these conditions, therefore we use only $\lambda$ is that case.

We can replace Eq.~(\ref{ec:lambda}) in Eq.~(\ref{ec:cdfqWeibulllambda}) obtaining
\begin{equation}
  I_{QW}(V;V_p,q,r,A,I_s)=\begin{cases} A\left(1-\left[1-\frac{(q-1)(r-1)}{(q-1)(r-1)-r} \left(\frac{V-V^*}{V_p-V^*}\right)^r\right]^{\frac{q-2}{q-1}} \right)+I_s &  ; r\neq 1  ,  \forall q\\
  A\left(1-\left[1- \left(\frac{V-V^*}{V_p-V^*}\right)\right]^{\frac{q-2}{q-1}} \right)+ I_s &  ;r= 1 ,  q\neq 1\\
  A\left(1-e^{\frac{V-V^*}{\lambda}}\right)+I_s &  ; r=1  ,  q= 1\\
  
  \end{cases}
  \label{ec:cdfqWeibullapp}
\end{equation}

\section*{Appendix C: Tsallis and Druyvesteyn EEPFs as particular cases of the q-Weibull EEPF}

This appendix is dedicated to link Eq.~(\ref{ec:cdfqWeibullapp}) with the known expressions of the non-extensive {\it{EEPF}}s, which are presented in Appendix B. First we write the second derivative of the probe current
\begin{equation}
  I^{\prime\prime}_{QW}\!(V)\!= 
  \begin{cases} 
   \frac{Ar(r-1)(q-2)}{(V-V^*)^{2-r}\lambda^r} \left(1- \left(\frac{V-V^*}{V_p-V^*}\right)^r\right) \left[1-(q-1)\left(\frac{V-V^*}{\lambda }\right)^r  \right]^{\frac{q}{1-q}} &  ;   r\neq1  , \forall q\\
\frac{A(q-2)}{\lambda^2}\left(1-(q-1)\frac{V-V^*}{\lambda}\right)^{\frac{q}{1-q}} &  ; r=1  ,  q\neq1
\\
\frac{-A}{\lambda^2}e^{\frac{V-V^*}{\lambda}} & ;r=1  , q=1

     \end{cases}
     \label{eq:iprimprimqwapp}
\end{equation}

Now, we consider the cases:

\subsection{Case $r=1$ and $q\neq 1$}

The {\it{EEPF}} can be obtained by the variable change $V$ to $V_p-V$, as is required by Druyvesteyn in Eq.~(\ref{ec:Druyvesteynf}), obtaining

\begin{equation}
    f_{QW}\!(V\!;V_p,q,r=1,A)\!= \frac{4A(q-2)}{e^3S\lambda^2}\sqrt{\frac{m}{2}}\left(1-(q-1)\frac{V_p-V^*-V}{\lambda}\right)^{\frac{q}{1-q}} \; , \; V>0 
\end{equation}
where $0<V<V_p-V^*$, 

\subsubsection{Silva's EEPF}
Defining $\overline{q}:= \frac{1}{q}$, $\varepsilon :=-e(V_p-V^*-V)$ , $kT :=e\lambda\overline{q}$   and ${B_q} := \frac{4A(1-2\overline{q})\overline{q}}{eS(kT)^2}\sqrt{\frac{m}{2}}$, we obtain
\begin{equation}
    f_{S}\!(\varepsilon)\!= {B_q} \left(1-(\overline{q}-1)\frac{\varepsilon}{kT}\right)^{\frac{1}{\overline{q}-1}} \; , \; 0<\varepsilon<-e(V_p-V^*) 
\end{equation}

which is the 3D EEPF obtained by Silva (Eq.~(\ref{ec:3DSilva})).

\subsubsection{Qiu and Liu's EEPF} 
Defining $q^\dagger:=2-q$ , $\varepsilon :=-e(V_p-V^*-V)$ , $kT := e \lambda$  and $A_q := \left(\frac{m}{2\pi kT}\right)^\frac{2}{3}\frac{-4q^\dagger A}{eS(kT)^2}\sqrt{\frac{m}{2}}$, we obtain
\begin{equation}
  f_{q^\dagger}(\varepsilon)=\left(\frac{m}{2\pi kT}\right)^\frac{3}{2} A_q \left(1-(q^\dagger-1)\frac{\varepsilon}{kT}\right)^{\frac{2-q^\dagger}{q^\dagger-1}} \; , \; 0<\varepsilon<-e(V_p-V^*) 
\end{equation}

which is the 3D distribution of electron energy obtained by Qiu and Liu (Eq.~(\ref{ec:3DQiu})).

\subsection{Case $r=1$ and $q=1$}

In this case, the {\it{EEPF}} result:
\begin{equation}
    f_{QW}\!(V\!;V_p,q=1,r=1,A)\!= -\frac{4A}{e^3S\lambda^2}\sqrt{\frac{m}{2}} e^{\frac{V_p-V^*-V}{\lambda}} \; , \; V>0 
\end{equation}
where $0<V<V_p-V^*$.

Defining  $\varepsilon :=-e(V_p-V^*-V)$ , $kT := e\lambda$  and $B := -\frac{4A}{eS(kT)^2}\sqrt{\frac{m}{2}}$, we obtain
\begin{equation}
    f_{BGS}\!(\varepsilon)\!= Be^{-\frac{\varepsilon}{kT}} \; , \; 0<\varepsilon<-e(V_p-V^*)
\end{equation}
which is the Boltzmann-Gibbs-Shannon EEPF.

\subsection{Case $r\neq 1$ and $q=1$}

The EEPF can be obtained from Eq.(\ref{eq:iprimprimqwapp}) by the variable change $V$ to $V_p-V$, as is required by Druyvesteyn in Eq.~(\ref{ec:Druyvesteynf}), obtaining

\begin{equation}
    f_{QW}\!(V\!;V_p,q=1,r,A)\!=\frac{-Ar(r-1)\left(1-\left(\frac{V_p-V^*-V}{V_p-V^*}\right)^r\right)}{(V_p-V^*-V)^{2-r}\lambda^r}  
e^{\left(\frac{V_p-V^*-V}{\lambda}\right)^r}    \; , \; V>0 
\end{equation}
where $0<V<V_p-V^*$. 
Defining $\varepsilon :=-e(V_p-V^*-V)$ and $kT := (-1)^{\frac{1}{r}}e\lambda$, we obtain
\begin{equation}
    f_{W}\!(\varepsilon)\!= \frac{e A_{\varepsilon}}{(kT)^r} e^{-\left(\frac{\varepsilon}{kT}\right)^r} \; ,
\end{equation}
which is the Weibull distribution or generalized EEPF. In the case $r=2$, it corresponds to the Druyvesteyn EEPF. For a recent review about these distributions, the reader is referred to \cite{Mouchtouris2018}, and the original works can be found in Refs. \cite{Ruzicka1970,Herrmann1971,Rundle1973}

\section*{Appendix D: Influence of the starting voltage (V*) }
This appendix analyzes the influence of the probe voltage point $V^*$, which is the more negative value of the {\bf{I-V}} curve. The point $V^*$ is somehow arbitrary because it depends on the starting point selected when measuring the {\bf{I-V}} Langmuir curve. In this Appendix we trim the experimental data from the more negative probe voltage values, then the more negative voltage $V^*$ changes accordingly, and we study the plasma parameters obtained by using the $q$-Weibull function with $I_s=I(V^*)$ as a fixed parameter. \\

\begin{table}[H]
\caption{Table that summarizes the fit parameters of the $q$-Weibull distribution for $h = 2.2$ mm  and the associated values of $T$ and $n$ calculated from the analytic expression of the second derivative. We selected different $V^*$ values and the experimental measurements $V<V^*$ have been trimmed. The fits have RMSE $\le 2\times 10^{-5}$. }
\label{tab:qweibullapp1}
\centering
\begin{tabular}{ c c c c c c c }
\br
$V^*\!\pm\! 0.001$ & $V_{p}\!\pm\! 0.005$ &  $q\!\pm\! 0.003$ & $r\!\pm\! 0.07$ & $A\!\pm\! 0.003$ & $n\!\pm\!0.05$ & $T\!\pm\! 150$ \\ 
 (V) & (V)  &  &  & (A) & ($10^{17}\!/$m$^3$) & (K) \\ [0.5ex]
\mr
-27.812 & -25.018 & 1.934 & 11.04  & 0.135 & 1.48 & 5180 \\
-27.763 & -25.014 & 1.932 & 10.64  & 0.134 & 1.49 & 5230 \\
-27.713 & -25.014 & 1.932 & 10.39  & 0.136 & 1.49 & 5230 \\
-27.663 & -25.013 & 1.932 & 10.13  & 0.137 & 1.49 & 5240 \\
-27.616 & -25.014 & 1.933 & 9.92   & 0.139 & 1.49 & 5220 \\
-27.563 & -25.014 & 1.933 & 9.67   & 0.141 & 1.49 & 5230 \\
-27.513 & -25.013 & 1.933 & 9.42   & 0.142 & 1.49 & 5230 \\
-27.462 & -25.015 & 1.935 & 9.25   & 0.146 & 1.48 & 5200 \\
-27.412 & -25.013 & 1.934 & 8.95   & 0.146 & 1.49 & 5220 \\
-27.362 & -25.014 & 1.935 & 8.73   & 0.149 & 1.49 & 5210 \\
-27.312 & -25.018 & 1.939 & 8.64   & 0.155 & 1.48 & 5140 \\
-27.262 & -25.013 & 1.936 & 8.22   & 0.153 & 1.49 & 5220 \\
-27.212 & -25.016 & 1.938 & 8.09   & 0.158 & 1.48 & 5170 \\
-27.162 & -25.016 & 1.939 & 7.83   & 0.161 & 1.48 & 5170 \\
-27.127 & -25.011 & 1.936 & 7.54   & 0.158 & 1.49 & 5230 \\
-27.077 & -25.011 & 1.936 & 7.29   & 0.161 & 1.49 & 5230 \\
-27.027 & -25.010 & 1.936 & 7.04   & 0.163 & 1.49 & 5240 \\
-26.977 & -25.010 & 1.936 & 6.79   & 0.165 & 1.49 & 5240 \\
-26.927 & -25.010 & 1.937 & 6.57   & 0.170 & 1.49 & 5230 \\
-26.877 & -25.010 & 1.936 & 6.29   & 0.170 & 1.49 & 5250 \\
-26.827 & -25.011 & 1.939 & 6.15   & 0.180 & 1.49 & 5190 \\
-26.777 & -25.010 & 1.937 & 5.85   & 0.180 & 1.49 & 5230 \\
-26.727 & -25.007 & 1.935 & 5.57   & 0.180 & 1.49 & 5250 \\
-26.677 & -25.010 & 1.937 & 5.38   & 0.187 & 1.49 & 5220 \\
-26.627 & -25.010 & 1.938 & 5.17   & 0.192 & 1.49 & 5200 \\
-26.577 & -25.010 & 1.938 & 4.94   & 0.197 & 1.49 & 5200 \\
\br
\end{tabular}
\end{table} 

We present two examples. The first example is shown in Table \ref{tab:qweibullapp1}, and depicts  the fitted plasma parameters with different $V^*$ starting from all the experimental data up to $V^*=V_f$. We only consider the height $h=2.2$ mm and the plasma parameters are obtained by the fitting procedure explained in Sec. \ref{sec:qweibull}. It is shown that plasma parameters $V_p$, $q$, $n$, and $T$ are practically independent of the $V^*$ value, besides  $r$ and $A$ differ depending on the different values of $V^*$. The $r$ parameter shows a decreasing behavior as $V^*$ increases, also the $r$ parameter allows the $q$-Weibull cdf to represent the concavity change which is present in $V = V_p$. 

The second example is shown in Table \ref{tab:qweibullapp2}, where we have considered $V^*=V_f$ for eight heights ranging from 2.2 to 9.4 mm. If we compare Table \ref{tab:qweibullapp2} with Table \ref{tab:qweibull}, we observe that the obtained parameters $V_p$, $q$, $n$, and $T$ are similar. To explain this result we consider two possible options:
\textbf{(1)} the $q$-Weibull function can represent the ion contributions with great precision; 
and \textbf{(2)} the second derivative of the $q$-Weibull function in the range from $V^*$ to $V_f$ (which corresponds with the ion contribution) is negligible, i.e. $I_p^{\prime\prime}\approx 0$ with $V$ in $(V^*,V_f)$. We have graphically observed that the $q$-Weibull function shows an almost horizontal behavior in this range of voltages from $V^*$ to $V_f$, which corresponds mainly to the ion current contribution, therefore the second option seems to be the correct one. We conclude that the $q$-Weibull function may have problems to represent the ion current contribution but it can represent the electron current contribution with great precision.

\begin{table}[H]
\caption{Table that summarizes the fit parameters  with $V^*=V_f$ and the associated values of $T$ and $n$ calculated from the analytic expression of the second derivative. The fits have RMSE $\le 2\times 10^{-5}$. }
\label{tab:qweibullapp2}
\centering
\begin{tabular}{ c c c c c c c }
\br
$h\!\pm\! 0.1$ & $V_{p}\!\pm\! 0.005$  & $q\!\pm\! 0.003$ & $r\!\pm\! 0.07$ & $A\!\pm\! 0.003$ & $n\!\pm\!0.05$ & $T\!\pm\! 150$ \\ 
(mm) & (V)  &  &  & (A) & ($10^{17}\!/$m$^3$) & (K) \\ [0.5ex]
\mr
2.2 & -25.008 & 1.940  & 4.94 & 0.197 & 1.59 & 5200 \\
3.4 & -25.134 & 1.940  & 3.97 & 0.237 & 1.52 & 5310 \\
4.3 & -25.396 & 1.970  & 6.73 & 0.263 & 1.26 & 3940 \\ 
5.1 & -25.520 & 1.970  & 6.76 & 0.206 & 1.15 & 3530 \\ 
6.2 & -25.671 & 1.964  & 6.13 & 0.154 & 1.06 & 3030 \\ 
7.3 & -25.721 & 1.950  & 6.51 & 0.098 & 0.97 & 2790 \\ 
8.2 & -25.850 & 1.950  & 7.01 & 0.087 & 0.88 & 2540 \\
9.4 & -26.013 & 1.950  & 6.36 & 0.077 & 0.76 & 2670 \\
\br
\end{tabular}
\end{table}

\typeout{}
\section*{References}
\bibliographystyle{unsrt}
\bibliography{Nonextensive}

\begin{thebibliography}{10}

\bibitem{Cherrington1982}
B.~E. Cherrington.
\newblock The use of electrostatic probes for plasma diagnostics.
\newblock {\em Plasma Chemistry and Plasma Processing}, 2(2):113--140, June
  1982.

\bibitem{Hershkowitz1989}
Noah Hershkowitz.
\newblock How langmuir probes work.
\newblock In {\em Plasma Diagnostics}, pages 113--183. Elsevier, 1989.

\bibitem{Merlino2007}
Robert~L. Merlino.
\newblock Understanding langmuir probe current-voltage characteristics.
\newblock {\em American Journal of Physics}, 75(12):1078--1085, December 2007.

\bibitem{Boyd1959}
R.~L.~F. Boyd and N.~D. Twiddy.
\newblock Electron energy distributions in plasmas. i.
\newblock {\em Proceedings of the Royal Society of London. Series A.
  Mathematical and Physical Sciences}, 250(1260):53--69, February 1959.

\bibitem{Demidov2002}
V.~I. Demidov, S.~V. Ratynskaia, and K.~Rypdal.
\newblock Electric probes for plasmas: The link between theory and instrument.
\newblock {\em Review of Scientific Instruments}, 73(10):3409--3439, October
  2002.

\bibitem{Abe2013}
Takumi Abe and Kohichiro Oyama.
\newblock Langmuir probe.
\newblock In {\em An Introduction to Space Instrumentation}, pages 63--75.
  {TERRAPUB}, 2013.

\bibitem{Bhattarai2017}
Shankar Bhattarai and Lekha~Nath Mishra.
\newblock Theoretical study of spherical langmuir probe in maxwellian plasma.
\newblock {\em International Journal of Physics}, 5(3):73--81, August 2017.

\bibitem{Woods1994}
R.~Claude Woods and Isaac~D. Sudit.
\newblock Theory of electron retardation by langmuir probes in anisotropic
  plasmas.
\newblock {\em Physical Review E}, 50(3):2222--2238, September 1994.

\bibitem{Knappmiller2006}
Scott Knappmiller, Scott Robertson, and Zoltan Sternovsky.
\newblock Method to find the electron distribution function from cylindrical
  probe data.
\newblock {\em Physical Review E}, 73(6), June 2006.

\bibitem{Godyak2011}
V~A Godyak and V~I Demidov.
\newblock Probe measurements of electron-energy distributions in plasmas: what
  can we measure and how can we achieve reliable results?
\newblock {\em Journal of Physics D: Applied Physics}, 44(23):233001, May 2011.

\bibitem{Druyvesteyn1930}
M.~J. Druyvesteyn.
\newblock Der niedervoltbogen.
\newblock {\em Zeitschrift f\"{u}r Physik}, 64(11-12):781--798, September 1930.

\bibitem{Waymouth1989}
John~F. Waymouth.
\newblock Plasma diagnostics in electric discharge light sources.
\newblock In {\em Plasma Diagnostics}, pages 47--111. Elsevier, 1989.

\bibitem{Hoegy1999}
Walter~R. Hoegy and Larry~H. Brace.
\newblock Use of langmuir probes in non-maxwellian space plasmas.
\newblock {\em Review of Scientific Instruments}, 70(7):3015--3024, July 1999.

\bibitem{Taccogna2016}
Francesco Taccogna and Giorgio Dilecce.
\newblock Non-equilibrium in low-temperature plasmas.
\newblock {\em The European Physical Journal D}, 70(11), November 2016.

\bibitem{Giono2017}
G~Giono, J~T Gudmundsson, N~Ivchenko, S~Mazouffre, K~Dannenmayer,
  D~Loub{\`{e}}re, L~Popelier, M~Merino, and G~Olent{\v{s}}enko.
\newblock Non-maxwellian electron energy probability functions in the plume of
  a {SPT}-100 hall thruster.
\newblock {\em Plasma Sources Science and Technology}, 27(1):015006, December
  2017.

\bibitem{Fiebrandt2017}
Marcel Fiebrandt, Moritz Oberberg, and Peter Awakowicz.
\newblock Comparison of langmuir probe and multipole resonance probe
  measurements in argon, hydrogen, nitrogen, and oxygen mixtures in a double
  {ICP} discharge.
\newblock {\em Journal of Applied Physics}, 122(1):013302, July 2017.

\bibitem{Tan1973}
W~P~S Tan.
\newblock Langmuir probe measurement of electron temperature in a druyvesteyn
  electron plasma.
\newblock {\em Journal of Physics D: Applied Physics}, 6(10):1206--1216, June
  1973.

\bibitem{Godyak1993}
V.~A. Godyak, R.~B. Piejak, and B.~M. Alexandrovich.
\newblock Probe diagnostics of non-maxwellian plasmas.
\newblock {\em Journal of Applied Physics}, 73(8):3657--3663, April 1993.

\bibitem{Gudmundsson2001}
J~T Gudmundsson.
\newblock On the effect of the electron energy distribution on the plasma
  parameters of an argon discharge: a global (volume-averaged) model study.
\newblock {\em Plasma Sources Science and Technology}, 10(1):76--81, January
  2001.

\bibitem{Seo2004}
Sang-Hun Seo and Hong-Young Chang.
\newblock Anomalous behaviors of plasma parameters in unbalanced direct-current
  magnetron discharge.
\newblock {\em Physics of Plasmas}, 11(7):3595--3601, July 2004.

\bibitem{Choe2009}
Jae-Myung Choe, Gon-Ho Kim, and Dai-Gyoung Kim.
\newblock The new langmuir probe analysis algorithm based on wavelet transforms
  to obtain electron energy distribution function of bi-maxwellian plasma.
\newblock {\em Journal of the Korean Physical Society}, 55(5(1)):1825--1835,
  November 2009.

\bibitem{Adams2017}
S.~F. Adams, J.~A. Miles, and V.~I. Demidov.
\newblock Non-maxwellian electron energy distribution function in a pulsed
  plasma modeled with dual effective temperatures.
\newblock {\em Physics of Plasmas}, 24(5):053508, May 2017.

\bibitem{Li2019}
Jinming Li, Ying Wang, Junjie Wei, Chengxun Yuan, Zhongxiang Zhou, Xiaoou Wang,
  and A.~A. Kudryavtsev.
\newblock Effects of non-maxwellian electron distribution function to the
  propagation coefficients of electromagnetic waves in plasma.
\newblock {\em {IEEE} Transactions on Plasma Science}, 47(1):100--103, January
  2019.

\bibitem{Sharma2018}
S.~Sharma, N.~Sirse, M.~M. Turner, and A.~R. Ellingboe.
\newblock Influence of excitation frequency on the metastable atoms and
  electron energy distribution function in a capacitively coupled argon
  discharge.
\newblock {\em Physics of Plasmas}, 25(6):063501, June 2018.

\bibitem{Tsallis1988}
Constantino Tsallis.
\newblock Possible generalization of boltzmann-gibbs statistics.
\newblock {\em Journal of Statistical Physics}, 52(1-2):479--487, July 1988.

\bibitem{Tsallis1995}
Constantino Tsallis.
\newblock Non-extensive thermostatistics: brief review and comments.
\newblock {\em Physica A: Statistical Mechanics and its Applications},
  221(1-3):277--290, November 1995.

\bibitem{Tsallis1997}
Constantino Tsallis.
\newblock L\'evy distributions.
\newblock {\em Physics World}, 10(7):42--46, Jul 1997.

\bibitem{Boghosian1996}
Bruce~M. Boghosian.
\newblock Thermodynamic description of the relaxation of two-dimensional
  turbulence using tsallis statistics.
\newblock {\em Physical Review E}, 53(5):4754--4763, May 1996.

\bibitem{Anteneodo1997}
Celia Anteneodo and Constantino Tsallis.
\newblock Two-dimensional turbulence in pure-electron plasma: A nonextensive
  thermostatistical description.
\newblock {\em Journal of Molecular Liquids}, 71(2-3):255--267, April 1997.

\bibitem{Silva1998}
R.~Silva, A.R. Plastino, and J.A.S. Lima.
\newblock A maxwellian path to the q-nonextensive velocity distribution
  function.
\newblock {\em Physics Letters A}, 249(5-6):401--408, December 1998.

\bibitem{Lima2000}
J.~A.~S. Lima, R.~Silva, and Janilo Santos.
\newblock Plasma oscillations and nonextensive statistics.
\newblock {\em Physical Review E}, 61(3):3260--3263, March 2000.

\bibitem{Du2004}
Jiulin Du.
\newblock Nonextensivity in nonequilibrium plasma systems with coulombian
  long-range interactions.
\newblock {\em Physics Letters A}, 329(4-5):262--267, August 2004.

\bibitem{Jiulin2007}
Du~Jiulin.
\newblock Nonextensivity and the power-law distributions for the systems with
  self-gravitating long-range interactions.
\newblock {\em Astrophysics and Space Science}, 312(1-2):47--55, September
  2007.

\bibitem{Budini2015}
Adri{\'{a}}n~A. Budini.
\newblock Extendedq-gaussian andq-exponential distributions from gamma random
  variables.
\newblock {\em Physical Review E}, 91(5), May 2015.

\bibitem{Qiu2018}
Hui-Bin Qiu and San-Qiu Liu.
\newblock Dispersion relation of longitudinal oscillation in relativistic
  plasmas with nonextensive distribution.
\newblock {\em Physics of Plasmas}, 25(10):102102, 2018.

\bibitem{Sun2020}
Futao Sun and Jiulin Du.
\newblock The collision frequency of electron-neutral-particle in the weakly
  ionized plasma with the power-law velocity distribution.
\newblock {\em Contributions to Plasma Physics}, 60(7):e201900183, March 2020.

\bibitem{ElBojaddaini2020}
Mohamed~El Bojaddaini and Hassan Chatei.
\newblock Ion source terms effect on collisional plasma sheath characteristics
  with non-extensively distributed electrons.
\newblock {\em The European Physical Journal Plus}, 135(8), August 2020.

\bibitem{Sharifian2014}
M.~M.~Sharifian, H. R. an~Shdarifinejad, Borhani Zarandi, and A.~R. Niknam.
\newblock Effect of q-non-extensive distribution of electrons on the plasma
  sheath floating potential.
\newblock {\em J. Plasma Physics}, 80(4):607--618, 2014.

\bibitem{Qiu2020}
Huibin Qiu, Zhenyu Zhou, Xingkun Peng, Xianyang Zhang, Yuqing Zhu, Yue Gao,
  Donghua Xiao, Haifeng Bao, Tianling Xu, Jia Zhang, and et~al.
\newblock Initial measurement of electron nonextensive parameter with electric
  probe.
\newblock {\em Physical Review E}, 101(4), Apr 2020.

\bibitem{Isola2009}
L~M Isola, B~J G{\'{o}}mez, and V~Guerra.
\newblock Determination of the electron temperature and density in the negative
  glow of a nitrogen pulsed discharge using optical emission spectroscopy.
\newblock {\em Journal of Physics D: Applied Physics}, 43(1):015202, December
  2009.

\bibitem{Huxley1974}
L.~G.~H. Huxley and R.~W. Crompton.
\newblock {\em The diffusion and drift of electrons in gases}.
\newblock New York, Wiley, 1974.

\bibitem{Ismail2015}
Ahmad~Fauzi Ismail, Kailash~Chandra Khulbe, and Takeshi Matsuura.
\newblock {\em Gas Separation Membranes}.
\newblock Springer International Publishing, 2015.

\bibitem{McDaniel1964}
Earl~W. McDaniel.
\newblock {\em Collision phenomena in ionized gases}.
\newblock New York, Wiley, 1964.

\bibitem{Biberman1987}
L.M. Biberman, Vorob'ev, V.S., and I.T. Yakubov.
\newblock {\em Kinetics of Nonequilibrium Low-Temperature Plasmas}.
\newblock Springer US, 1987.

\bibitem{Faudot2019}
E.~Faudot, J.~Ledig, J.~Moritz, S.~Heuraux, N.~Lemoine, and S.~Devaux.
\newblock Experimental measurements of the {RF} sheath thickness with a
  cylindrical langmuir probe.
\newblock {\em Physics of Plasmas}, 26(8):083503, August 2019.

\bibitem{Langmuir1926}
H.~M. Mott-Smith and Irving Langmuir.
\newblock The theory of collectors in gaseous discharges.
\newblock {\em Phys. Rev.}, 28:727--763, Oct 1926.

\bibitem{Godyak1992RF}
V~A Godyak, R~B Piejak, and B~M Alexandrovich.
\newblock Measurement of electron energy distribution in low-pressure {RF}
  discharges.
\newblock {\em Plasma Sources Science and Technology}, 1(1):36--58, March 1992.

\bibitem{Lieberman}
Michael~A. Lieberman and Allan~J. Lichtenberg.
\newblock {\em Principles of Plasma Discharges and Materials Processing}.
\newblock John Wiley Sons, Inc., Jan 2005.

\bibitem{Godyak1990}
V.~A. Godyak.
\newblock Measuring {EEDF} in gas discharge plasmas.
\newblock In {\em Plasma-Surface Interactions and Processing of Materials},
  pages 95--134. Springer Netherlands, 1990.

\bibitem{Magnus2008}
F.~Magnus and J.~T. Gudmundsson.
\newblock Digital smoothing of the langmuir probe i-v characteristic.
\newblock {\em Review of Scientific Instruments}, 79(7):073503, July 2008.

\bibitem{Savitzky-Golay}
Abraham Savitzky and M.~J.~E. Golay.
\newblock {Smoothing and Differentiation of Data by Simplified Least Squares
  Procedures.}
\newblock {\em Anal. Chem.}, 36(8):1627--1639, July 1964.

\bibitem{Roh2015}
Hyun-Joon Roh, Nam-Kyun Kim, Sangwon Ryu, Seolhye Park, Seok-Hwan Lee,
  Sung-Ryul Huh, and Gon-Ho Kim.
\newblock Determination of electron energy probability function in
  low-temperature plasmas from current {\textendash} voltage characteristics of
  two langmuir probes filtered by savitzky{\textendash}golay and blackman
  window methods.
\newblock {\em Current Applied Physics}, 15(10):1173--1183, October 2015.

\bibitem{Godyak2021}
Valery Godyak.
\newblock Rf discharge diagnostics: Some problems and their resolution.
\newblock {\em Journal of Applied Physics}, 129(4):041101, 2021.

\bibitem{Godyak1990PhRevLett}
V.~A. Godyak and R.~B. Piejak.
\newblock Abnormally low electron energy and heating-mode transition in a
  low-pressure argon rf discharge at 13.56 mhz.
\newblock {\em Phys. Rev. Lett.}, 65:996--999, Aug 1990.

\bibitem{Lipschultz1986}
B.~Lipschultz, I.~Hutchinson, B.~LaBombard, and A.~Wan.
\newblock Electric probes in plasmas.
\newblock {\em Journal of Vacuum Science {\&} Technology A: Vacuum, Surfaces,
  and Films}, 4(3):1810--1816, May 1986.

\bibitem{PicoliJr2009}
S.~Picoli Jr., R.~S. Mendes, L.~C. Malacarne, and R.~P.~B. Santos.
\newblock q-distributions in complex systems: a brief review.
\newblock {\em Brazilian Journal of Physics}, 39(2a):468--474, August 2009.

\bibitem{Zhang2018}
Fode Zhang, Hon Keung~Tony Ng, and Yimin Shi.
\newblock On alternative q-weibull and q-extreme value distributions:
  Properties and applications.
\newblock {\em Physica A: Statistical Mechanics and its Applications},
  490:1171--1190, January 2018.

\bibitem{Abbas2020}
Nasir Abbas.
\newblock On examining complex systems using the q-weibull distribution in
  classical and bayesian paradigms.
\newblock {\em Journal of Statistical Theory and Applications}, Vol.
  19(3):368--382, Sep 2020.

\bibitem{Masi2005}
Marco Masi.
\newblock A step beyond tsallis and r{\'{e}}nyi entropies.
\newblock {\em Physics Letters A}, 338(3-5):217--224, May 2005.

\bibitem{Jizba2017}
Petr Jizba and Jan Korbel.
\newblock On the uniqueness theorem for pseudo-additive entropies.
\newblock {\em Entropy}, 19(11):605, November 2017.

\bibitem{Lins2018}
Isis Lins, M{\'{a}}rcio Moura, Enrique Droguett, and Tha{\'{\i}}s Corr{\^{e}}a.
\newblock Combining generalized renewal processes with non-extensive
  entropy-based q-distributions for reliability applications.
\newblock {\em Entropy}, 20(4):223, March 2018.

\bibitem{Frank2000}
T.D. Frank and A.~Daffertshofer.
\newblock Exact time-dependent solutions of the renyi fokker–planck equation
  and the fokker–planck equations related to the entropies proposed by sharma
  and mittal.
\newblock {\em Physica A: Statistical Mechanics and its Applications},
  285(3):351--366, 2000.

\bibitem{Xu2017}
Meng Xu, Enrique~L{\'{o}}pez Droguett, Isis~Didier Lins, and M{\'{a}}rcio das
  Chagas~Moura.
\newblock On the q-weibull distribution for reliability applications: An
  adaptive hybrid artificial bee colony algorithm for parameter estimation.
\newblock {\em Reliability Engineering {\&} System Safety}, 158:93--105,
  February 2017.

\bibitem{Li2020}
Chen Li, Thorsten Hofmann, Klaus Edinger, Valery Godyak, and Gottlieb~S.
  Oehrlein.
\newblock Etching of si3n4 induced by electron beam plasma from hollow cathode
  plasma in a downstream reactive environment.
\newblock {\em Journal of Vacuum Science {\&} Technology B}, 38(3):032208, May
  2020.

\bibitem{Longhurst1981}
G.~Longhurst.
\newblock A method for obtaining electron energy density functions from
  langmuir probe data using a card-programable calculator.
\newblock In {\em 15th International Electric Propulsion}. American Institute
  of Aeronautics and Astronautics, April 1981.

\bibitem{Winkler1973ZurMBI}
R.~Winkler and S.~Pfau.
\newblock Zur mikrophysikalischen beschreibung des schwachionisierten
  stickstoffmolek{\"u}lplasmas der positiven s{\"a}ule von glimmentladungen. i.
  berechnung der geschwindigkeitsverteilungsfunktion der elektronen im
  molekularem stickstoffplasma und vergleich mit dem experiment.
\newblock {\em Contributions To Plasma Physics}, 13:273--295, 1973.

\bibitem{Winkler1977ZurMBII}
R.~Winkler and S.~Pfau.
\newblock Zur mikrophysikalischen beschreibung des schwachionisierten
  s{\"a}ulenplasmas von glimmentladungen in stickstoff‐neon‐gemischen. ii.
  aus der kinetischen gleichung berechnete transportgr{\"o}{\ss}en,
  sto{\ss}frequenzen und energieverlustraten der elektronen des mischplasmas.
\newblock {\em Contributions To Plasma Physics}, 17:317--336, 1977.

\bibitem{Pfau1977ZurMBIII}
S.~Pfau and R.~Winkler.
\newblock Zur mikrophysikalischen beschreibung des schwachionisierten
  s{\"a}ulenplasmas von glimmentladungen in stickstoff‐neon‐gemischen iii.
  bilanzierung des s{\"a}ulenplasmas und berechnung seiner inneren parameter.
\newblock {\em Contributions To Plasma Physics}, 17:397--417, 1977.

\bibitem{Hippler2008}
R.~Hippler, H.~Kersten, M.~Schmidt, and K.~H. Schoenbach.
\newblock {\em Low Temperature Plasmas: Fundamentals, Technologies and
  Techniques}.
\newblock Wiley-VCH, March 2008.

\bibitem{Olson2010}
J.~Olson, N.~Brenning, J.-E. Wahlund, and H.~Gunell.
\newblock On the interpretation of langmuir probe data inside a spacecraft
  sheath.
\newblock {\em Review of Scientific Instruments}, 81(10):105106, October 2010.

\bibitem{Samaniego2019}
Joseph~I. Samaniego and Xu~Wang.
\newblock Retrieving true plasma characteristics from langmuir probes immersed
  in the spacecraft sheath: The double hemispherical probe technique.
\newblock {\em Journal of Geophysical Research: Space Physics},
  124(12):9847--9856, December 2019.

\bibitem{Nie2018}
L.~Nie, M.~Xu, R.~Ke, B.D. Yuan, Y.F. Wu, J.~Cheng, T.~Lan, Y.~Yu, R.J. Hong,
  D.~Guo, L.~Ting, Y.B. Dong, Y.P. Zhang, X.M. Song, W.L. Zhong, Z.H. Wang,
  A.P. Sun, J.Q. Xu, W.~Chen, L.W. Yan, X.L. Zou, X.R. Duan, and et~al.
\newblock Experimental evaluation of langmuir probe sheath potential
  coefficient on the {HL}-2a tokamak.
\newblock {\em Nuclear Fusion}, 58(3):036021, February 2018.

\bibitem{Denysenko2015}
I.~B. Denysenko, H.~Kersten, and N.~A. Azarenkov.
\newblock Electron energy distribution in a dusty plasma: Analytical approach.
\newblock {\em Physical Review E}, 92(3), September 2015.

\bibitem{Kortshagen1994}
U.~Kortshagen.
\newblock Experimental evidence on the nonlocality of the electron distribution
  function.
\newblock {\em Physical Review E}, 49(5):4369--4380, May 1994.

\bibitem{Maresca2002}
Antonio Maresca, Konstantin Orlov, and Uwe Kortshagen.
\newblock Experimental study of diffusive cooling of electrons in a pulsed
  inductively coupled plasma.
\newblock {\em Physical Review E}, 65(5), May 2002.

\bibitem{Denysenko2004}
I.~Denysenko, M.~Y. Yu, K.~Ostrikov, and A.~Smolyakov.
\newblock Spatially averaged model of complex-plasma discharge with
  self-consistent electron energy distribution.
\newblock {\em Physical Review E}, 70(4), October 2004.

\bibitem{Takahashi2011}
Kazunori Takahashi, Christine Charles, Rod~W. Boswell, and Tamiya Fujiwara.
\newblock Electron energy distribution of a current-free double layer:
  Druyvesteyn theory and experiments.
\newblock {\em Physical Review Letters}, 107(3), July 2011.

\bibitem{Boswell2015}
Rod~W. Boswell, Kazunori Takahashi, Christine Charles, and Igor~D. Kaganovich.
\newblock Non-local electron energy probability function in a plasma expanding
  along a magnetic nozzle.
\newblock {\em Frontiers in Physics}, 3, March 2015.

\bibitem{Darian2019}
D~Darian, S~Marholm, M~Mortensen, and W~J Miloch.
\newblock Theory and simulations of spherical and cylindrical langmuir probes
  in non-maxwellian plasmas.
\newblock {\em Plasma Physics and Controlled Fusion}, 61(8):085025, June 2019.

\bibitem{Mouchtouris2018}
S.~Mouchtouris and G.~Kokkoris.
\newblock A generalized electron energy probability function for inductively
  coupled plasmas under conditions of nonlocal electron kinetics.
\newblock {\em Journal of Applied Physics}, 123(2):023301, January 2018.

\bibitem{Ruzicka1970}
T.~R{\r{u}}{\v{z}}i{\v{c}}ka, A.~Rutscher, and S.~Pfau.
\newblock Die transportkoeffizienten der elektronen in den edelgasen bei
  mittleren reduzierten elektrischen feldst\"{a}rken (teil {II}) (e/p0 $\asymp$
  0, 1{\ldots}10v/cm torr) thermodiffusions- und energietransportkoeffizienten
  der elektronen.
\newblock {\em Annalen der Physik}, 479(3-4):124--135, 1970.

\bibitem{Herrmann1971}
D.~Herrmann, A.~Rutscher, and S.~Pfau.
\newblock Radiale \"{A}nderungen der energieverteilungsfunktion der elektronen
  im plasma der positiven s\"{a}ule elektrischer entladungen.
\newblock {\em Beitr\"{a}ge aus der Plasmaphysik}, 11(1):75--84, 1971.

\bibitem{Rundle1973}
H.~W. Rundle, D.~R. Clark, and J.~M. Deckers.
\newblock Electron energy distribution functions in an o2 glow discharge.
\newblock {\em Canadian Journal of Physics}, 51(2):144--148, 1973.

\end{thebibliography}

\end{document}